\documentclass{article}
\usepackage{graphicx} 
\usepackage{amsfonts}
\usepackage{amsmath}
\usepackage{amsthm}
\usepackage{amssymb}
\usepackage{geometry}
\usepackage{enumitem}
\usepackage{xcolor}
\newtheorem{definition}{Definition}

\usepackage{url}
\usepackage{hyperref}

\usepackage{authblk}
\usepackage{float}

\title{Post-Quantum Homomorphic Encryption: A Case for Code-Based Alternatives}
\author[1]{Siddhartha Siddhiprada Bhoi}
\author[1]{Arathi Arakala}
\author[1]{Amy Beth Corman}
\author[1]{Asha Rao}

\affil[1]{School of Computing Technologies, RMIT University, Australia}
\affil{\vspace{0.5em} Emails: \textit{siddhartha.bhoi@student.rmit.edu.au, arathi.arakala@rmit.edu.au, amy.corman@rmit.edu.au, asha.rao@rmit.edu.au}}

\date{}
\begin{document}

\maketitle

\begin{abstract}
Homomorphic Encryption (HE) allows secure and privacy-protected computation on encrypted data without the need to decrypt it. Since Shor's algorithm rendered prime factorisation and discrete logarithm-based ciphers insecure with quantum computations, researchers have been working on building post-quantum homomorphic encryption (PQHE) algorithms. Most of the current PQHE algorithms are secured by Lattice-based problems and there have been limited 
attempts to build ciphers based on error-correcting code-based problems. This review presents an overview of the current approaches to building PQHE schemes and justifies code-based encryption as a novel way to diversify post-quantum algorithms. We present the mathematical underpinnings of existing code-based cryptographic frameworks and their security and efficiency guarantees. We compare lattice-based and code-based homomorphic encryption solutions identifying challenges that have inhibited the progress of code-based schemes. We finally propose five new research directions to advance post-quantum code-based homomorphic encryption.
\end{abstract}


\section{Introduction}
Homomorphic encryption (HE) allows computations to be performed directly on encrypted data without decryption, enabling privacy-preserving computation in applications such as secure cloud computing, medical data analysis, and federated learning \cite{Aloufi2021}. HE ensures the confidentiality of sensitive data while still being usable for computation, making it a critical tool in modern cryptography.
Homomorphic properties were first observed in classical public-key cryptosystems by Rivest et al \cite{rivest1978data} where certain operations like multiplication could be performed homomorphically on the ciphertexts. Although the scheme was broken a decade later by Brickell and Yacobi \cite{brickell1988privacy}, it initiated the research on homomorphic encryption (HE) schemes in the cryptography community. 

Many early homomorphic encryption schemes were based on number-theoretic assumptions of NP-hardness. However, with the advent of quantum computing, classical cryptographic systems face significant threats. Quantum algorithms, such as Shor's and Grover's algorithms, can efficiently solve problems such as integer factorization and discrete logarithms, thus undermining the security foundations of widely used public-key schemes such as RSA and ECC \cite{shor1999polynomial}. As many early homomorphic encryption (HE) constructions rely on these same number-theoretic assumptions, the potential for quantum adversaries to break the underlying hard problems necessitated a comprehensive re‐examination of HE for a post‐quantum era. In response, research has increasingly focused on developing post-quantum homomorphic encryption (PQHE) schemes built on hard mathematical problems believed to be resistant to quantum attacks.

In 2009, Gentry \cite{gentry2009fully} proposed the first FHE scheme in his Ph.D. thesis, which allowed both addition and multiplication operations to be performed on ciphertexts without observation of the plaintexts. The scheme used NP-hard problems based on lattice-theory. Subsequently, lattice-based cryptography emerged as a leading candidate to construct PQHE schemes. Cryptographic primitives based on the Learning With Errors (LWE) problem and its ring variant (RLWE) also offer robust security guarantees as no efficient quantum algorithm is currently known for solving these problems — even approximately \cite{li2022tutorial}. These lattice‐based approaches underpin modern schemes such as the Brakerski–Fan–Vercauteren (BFV) and Cheon–Kim–Kim–Song (CKKS) constructions, which have been reevaluated and adapted to meet post‐quantum security requirements \cite{brakerski2014leveled}. 
Parallel to these theoretical advancements, practical frameworks such as PALISADE \cite{palisade} and its successor OpenFHE \cite{polyakov2022openfhe} have integrated lattice‐based HE schemes optimized for post‐quantum security. These efforts underscore the practical drive to transition existing systems to quantum‐resistant alternatives while addressing efficiency challenges, such as managing noise growth during computations and optimizing bootstrapping procedures, to maintain the homomorphic properties essential for real-world applications \cite{chen2023homomorphic}. Although most of the focus of HE schemes is still based on lattice-theory, in 2024 Chen proposed new techniques involving Quantum Fourier Transforms that claimed to solve LWE problems and their reductions to other lattice-based hard problems in polynomial time \cite{chen2024quantum}. While the validity of the claim is questionable \cite{Smart_LWE}, it raised concerns about relying completely on the security of lattice-based cryptographic algorithms in the post-quantum era.

A complementary, though less explored, approach to developing post-quantum homomorphic encryption (PQHE) schemes involves NP-hard problems related to error-correcting codes. Code-based cryptography has emerged as a promising option for post-quantum homomorphic encryption due to its inherent resistance to quantum attacks and potential efficiency advantages. Here, security assumptions are based on NP-hard problems, such as the Syndrome Decoding Problem (SDP) and the Binary Goppa Code Distinguishing Problem \cite{berlekamp2003inherent}, both of which are resilient to known quantum algorithms. Unlike lattice-based schemes, which depend on complex algebraic structures and intricate noise management techniques, code-based homomorphic encryption utilizes well-established error-correcting codes. This makes implementation simpler and allows for alternative security assumptions. The specific trade-offs between security, efficiency, and implementation complexity highlight the importance of exploring code-based methods for future cryptographic systems, particularly in a post-quantum landscape.


This review aims to be a resource for researchers wanting to advance PQHE using the less-studied code-based cryptography. It brings together the mathematical underpinnings of code-based HE schemes with the intention of deepening understanding of the NP-hard problems in this domain as well as presenting avenues to kickstart new research in the development of code-based PQFHE. The key contributions of this work are:
\begin{enumerate}
    \item \textbf{Insights from reviews of existing Homomorphic Encryption Schemes: } Identifies the lop-sided focus of current FHE schemes on lattice-based NP-hard problems. This demonstrates a clear weakness in PQFHE schemes if quantum lattice reduction algorithms are found. 
    
    \item \textbf{Systematic Review of Code-based Homomorphic Encryption Schemes: } Provides a detailed analysis of code-based Homomorphic encryption schemes and their security guarantees.
    \item \textbf{Comparative Analysis of Code-Based with Lattice-based Homomorphic Encryption schemes:} Presents a detailed comparison of code-based HE with lattice-based HE and identifies the reasons code-based HE is not well studied.
    \item \textbf{New directions for Code-based Homomorphic Encryption in the Post-Quantum Era:} Outlines five clear research directions to advance the development of code-based PQ-FHE.
    
\end{enumerate}

We will first analyse the existing reviews of homomorphic encryption schemes, and illustrate the difference of this review. In Section \ref{Sec-Background}  we provide a background on homomorphic encryption schemes categorising them into partially homomorphic, somewhat homomorphic and fully homomorphic. Then in Section \ref{Sec:CBC} we present the details of code-based cryptography, while in Section \ref{Sec:CBHES} we detail code-based homomorphic schemes, giving a comprehensive analysis of the strengths and weaknesses of each scheme and comparing them to lattice-based schemes.  Finally, we present the conclusion along with the challenges in implementing code-based PQHE  and a list of future research opportunities.

\section{Existing reviews of homomorphic encryption schemes}
\begin{figure}
    \centering
    \includegraphics[width=0.5\linewidth]{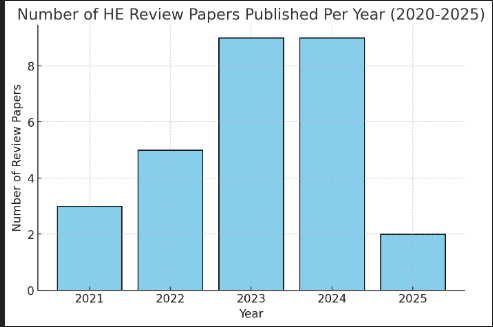}
    \caption{Distribution of review papers on Homomomorphic Encryption published in last 5 years by year of publication.}
    \label{fig:year_wise}
\end{figure}

\begin{figure}
    \centering
    \includegraphics[width=0.5\linewidth]{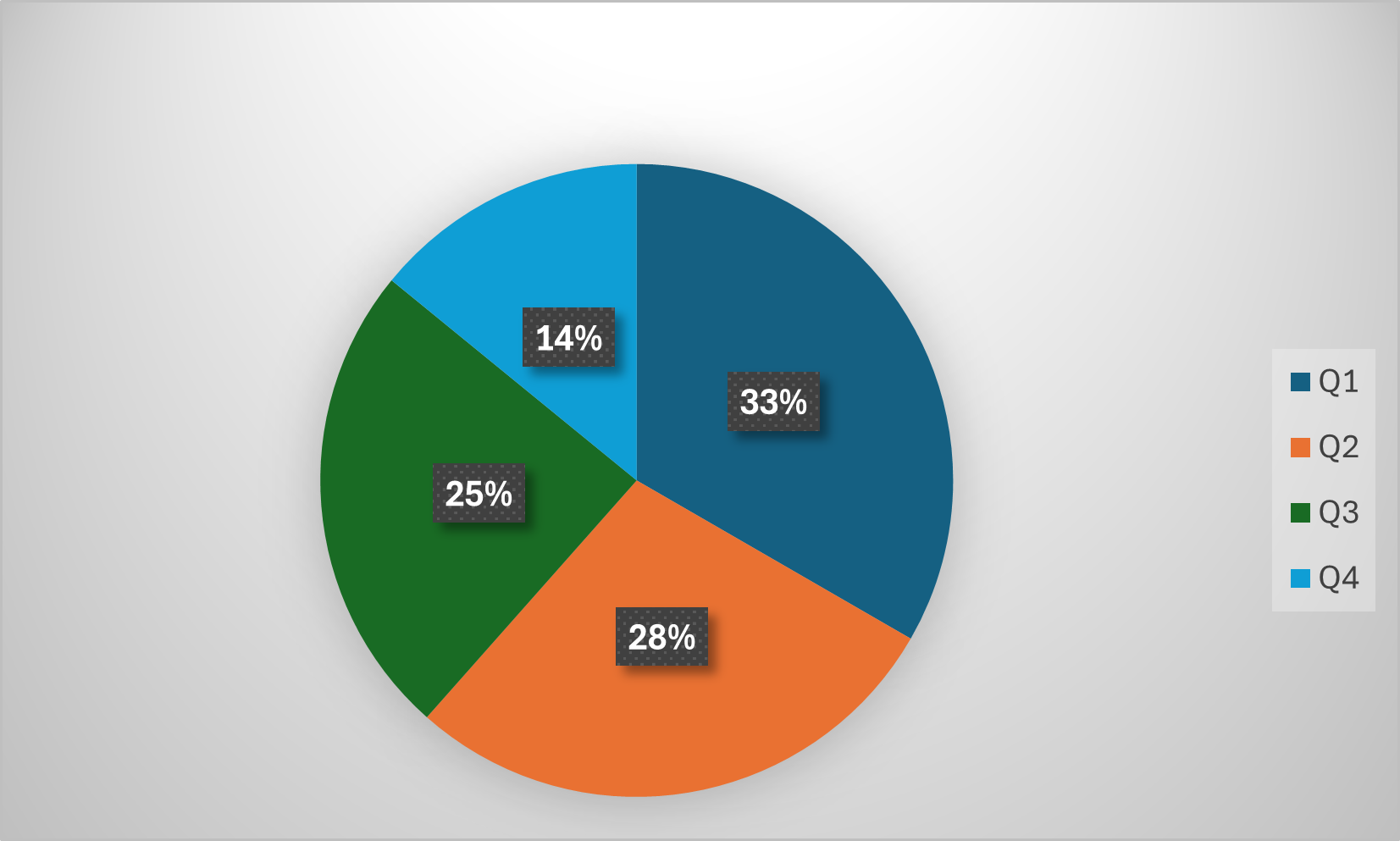}
    \caption{Scimago\cite{sci} journal quartile distribution of review papers on Homomorphic Encryption in 5 years}
    \label{fig:quartile_wise}
\end{figure}

To maximise the relevance of the current review, we analysed existing reviews on HE schemes. Scopus and IEEE databases were chosen as the source of our search process. On both databases we used the search terms  ``Homomorphic Encryption" and ``Privacy-Preserving Encryption", limited our results to only review articles and did not put limits on the year of publication. We found 108 articles, with the first review published in 2010. We found two duplicates across the database and they were removed. We screened the title and abstracts for relevance and excluded one article as it did not focus on HE or its applications. We then filtered out all articles published in the year 2020 and prior. This gave us 78 review articles on HE published in the last 5 years. Figure \ref{fig:year_wise} illustrates the distribution of published articles over the last 5 years. The trend indicates a surge of research since 2023 driven by advancements in HE applications. Figure \ref{fig:quartile_wise} presents the quality ratings of the journals where the review articles were published. It can be seen that only $33\%$ (26 articles) of the published reviews are in high quality Q1 journals. Most of the Q1 articles focussed on applications of HE in various fields rather than on the HE schemes and their theory. There were only four theoretical, non-application-based review articles and these were selected for further analysis.

Alaya et al. \cite{trends_challenges} provide an overview of homomorphic encryption (HE) systems, highlighting trends in efficiency, security, and usability, while addressing challenges such as computational overhead and scalability. Aloufi et al. \cite{Aloufi2021} investigate multi-party computations and secure collaborative analytics, discussing various schemes, security guarantees, and practical applications. Gorantala et al. \cite{unlocking_fhe} explore advancements in fully homomorphic encryption (FHE) and its real-world feasibility, focusing on performance improvements and necessary breakthroughs for secure cloud computing. Dhiman et al. \cite{he_library_review} evaluate software libraries and tools for homomorphic encryption, comparing their performance and usability to assist researchers in tool selection.

Table \ref{tab:he_coverage} compares the range of topics covered by these four articles. We assessed whether the reviews provided an overview of HE systems, discussed multi-key HE, included research on Fully Homomorphic Encryption (FHE), presented libraries or frameworks to practically apply HE, provided description of HE toolkits, indicated techniques to speed up HE algorithms including refresh key generation, identified existing and foreseeable challenges in the domain of HE and presented future trends in HE. We also investigated if any of these reviews included HE schemes that used code-based cryptography. As shown in the last row of Table \ref{tab:he_coverage}, none of the review articles mention code-based HE or discussed the advantages of using such approaches in the post-quantum era. The current review starts to address this deficiency in the existing literature by providing a first step towards developing an alternative code-based approach for HE in the post-quantum era.


\begin{table}[H]
\caption{Coverage of Homomorphic Encryption Topics in Previous Reviews}
\centering
\begin{tabular}{|p{5.5cm}|p{1.5cm}|p{1.5cm}|p{1.5cm}|p{1.5cm}|}
\hline
\textbf{Aspect} & \textbf{Alaya et.al. \cite{trends_challenges}} & \textbf{Aloufi et.al. \cite{Aloufi2021}} & \textbf{Gorantala et.al. \cite{unlocking_fhe}} & \textbf{Dhiman et.al.\cite{he_library_review}} \\
\hline
Overview of HE systems & \checkmark & \checkmark & \checkmark & \checkmark \\
\hline
Multi-key HE & $\times$ & \checkmark & $\times$ & $\times$ \\
\hline
Fully Homomorphic Encryption (FHE) & \checkmark & \checkmark & \checkmark & \checkmark \\
\hline
HE Libraries/ Frameworks & $\times$ & $\times$ & $\times$ & \checkmark \\
\hline
HE Toolkits & $\times$ & $\times$ & $\times$ & \checkmark \\
\hline
HE Accelerators & $\times$ & $\times$ & $\times$ & \checkmark \\
\hline
Challenges in HE & \checkmark & $\times$ & \checkmark & $\times$ \\
\hline
Future Trends & \checkmark & $\times$ & \checkmark & $\times$ \\
\hline
 \textbf{Code-based HE} & \textbf{$\times$} & \textbf{$\times$} & \textbf{$\times$} & \textbf{$\times$} \\
\hline
\end{tabular}

\label{tab:he_coverage}
\end{table}

\section{Background on Homomorphic Encryption}\label{Sec-Background}
This section provides an overview of homomorphic encryption (HE), explaining its basic principles and levels of capability. Homomorphic encryption allows computations on encrypted data without decryption. Each of the different types: partially homomorphic encryption (PHE), somewhat homomorphic encryption (SWHE), and fully homomorphic encryption (FHE), offers varying degrees of flexibility and efficiency. This background helps in understanding the advancements and challenges in HE research.

\subsection{Homomorphic Encryption: Definitions} 

Homomorphic encryption (HE) is a cryptographic technique that allows computations to be performed on encrypted data without the need for decryption. This property enables the processing and analysis of sensitive information while preserving its confidentiality and integrity. 

\subsubsection{Formal definition of Homomorphic Encryption}
Formally, a homomorphic encryption scheme can be defined as a tuple of probabilistic polynomial-time (PPT) algorithms, denoted as \( \text{HE} = (\text{KeyGen}, \text{Enc}, \text{Eval}, \text{Dec}) \) \cite{Aloufi2021}. Each algorithm is described in detail below:

\begin{itemize}
    \item \textbf{Key Generation (KeyGen):}
       \( \text{HE.KeyGen}(1^\lambda) \rightarrow (pk, sk, ek) \): 
        Given a security parameter \( \lambda \) that determines the level of security, the key generation algorithm outputs a public key \( pk \), a secret key \( sk \), and an evaluation key \( ek \). The public key is used for encryption, the secret key for decryption, and the evaluation key for performing homomorphic operations.

    \item \textbf{Encryption (Enc):}
    \( \text{HE.Enc}(pk, m) \rightarrow c \): 
        Given a message \( m \) and the public key \( pk \), the encryption algorithm outputs a ciphertext \( c \). The ciphertext \( c \) is an encrypted version of the message \( m \), which can be processed homomorphically without revealing the underlying plaintext.

    \item \textbf{Evaluation (Eval):}
    \( \text{HE.Eval}(ek, f, c, c') \rightarrow c_{\text{eval}} \): 
        Given two ciphertexts \( c \) and \( c' \), an evaluation key \( ek \), and a homomorphic function \( f \), the evaluation algorithm outputs an evaluated ciphertext \( c_{\text{eval}} = f(c, c') \). The evaluation key \( ek \) is used to enable homomorphic operations, and it plays a crucial role in the bootstrapping process, which refreshes the ciphertext to allow for further computations.

    \item \textbf{Decryption (Dec):}
     \( \text{HE.Dec}(sk, c) \rightarrow m \): 
        Given a ciphertext \( c \) encrypted under the public key \( pk \), the decryption algorithm outputs the original message \( m \) using the corresponding secret key \( sk \). This ensures that only the holder of the secret key can access the decrypted data.

\end{itemize}

\begin{figure}
    \centering
    \includegraphics[width=1\linewidth]{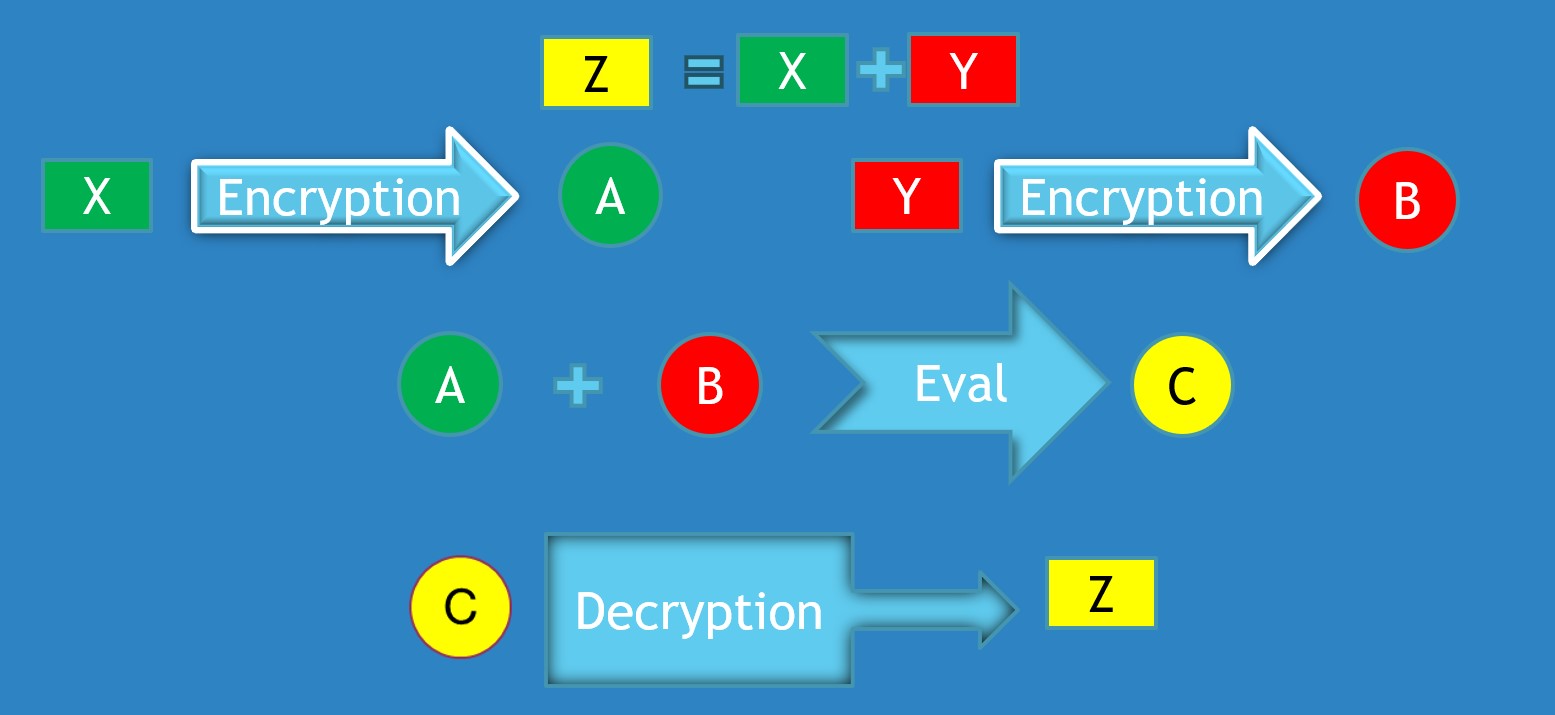}
    \caption{Illustration of the general Homomorphic Encryption process. Here $X, Y$ are plaintexts, $A, B$ are encryptions of $X, Y$ respectively, $C$ is the homomorphic evaluation of $A, B$, while $Z$ is the decrypted value of $C$.}
    \label{fig:HE_Basics}
\end{figure}

Figure \ref{fig:HE_Basics} provides a conceptual overview of how homomorphic encryption 
allows computations to be performed directly on encrypted data. The process 
happens in four main stages:

\begin{enumerate}
    \item \textbf{Plaintext Inputs (\(X, Y\)):} Two plaintext values, labeled  \(X\) (green) and \(Y\) (red), are shown in Figure~\ref{fig:HE_Basics}. In this example, the goal is to compute the sum \(Z = X + Y\) securely.

    \item \textbf{Encryption (\(A, B\)):} Each plaintext is independently encrypted, producing ciphertexts \(A\) (encrypted form of \(X\)) and \(B\) 
    (encrypted form of \(Y\)). These ciphertexts are represented by the green  and red circles, respectively.

    \item \textbf{Homomorphic Evaluation (\(C\)):} The ciphertexts \(A\) and  \(B\) are processed by a homomorphic evaluation function (indicated by the arrow labeled ``Eval''). Because the scheme is homomorphic, the evaluation computes the sum of the underlying plaintexts---i.e., it effectively adds \(X\) and \(Y\)---without ever decrypting them. This step yields a new ciphertext \(C\).

    \item \textbf{Decryption (\(Z\)):} Finally, decrypting \(C\) recovers the result \(Z\), which matches the sum \(X + Y\). In other words, \(\mathrm{Dec}(C) = Z = X + Y\).
    
\end{enumerate}

\subsubsection{Mathematical definition of Homomorphic Encryption}

\noindent Mathematically, a homomorphic encryption (HE) scheme is defined as follows: 
\vspace{-1mm}
\begin{definition}
    A homomorphic encryption (HE) scheme is an encryption scheme that satisfies the following property: 
For all ciphertexts \( C_1, C_2 \in \mathcal{C} \), all plaintexts \( M_1, M_2 \in \mathcal{P} \), and for any key \( K \), if 
\[
C_1 = \operatorname{Enc}(M_1) \quad \text{and} \quad C_2 = \operatorname{Enc}(M_2),
\]
then 
\[
\operatorname{Dec}\bigl( C_1 \circ C_2 \bigr) = M_1 \odot M_2. \tag{1}
\]
Here, \(\circ\) and \(\odot\) denote the group operations in the ciphertext space \(\mathcal{C}\) and the plaintext space \(\mathcal{P}\), respectively.
\end{definition}

\subsection{Levels of Homomorphism}
HE schemes are often classified into three levels based on the types and number of homomorphic operations that can be supported. These are:
\begin{itemize}
    \item \textbf{Partially Homomorphic Encryption (PHE)}: Supports only one operation (e.g., addition or multiplication).
    \item \textbf{Somewhat Homomorphic Encryption (SHE)}: Supports all operations, but has limitations on the number of operations of a certain type.
    \item \textbf{Fully Homomorphic Encryption (FHE)}: Enables unlimited arbitrary computations on encrypted data.
\end{itemize}


Table \ref{tab:homomorphism} compares the salient features of the different levels of homomorphic encryption.

\begin{table}[h!]
\caption{ Features of Partial HE, Somewhat HE, and Fully HE.}
\begin{tabular}{|p{1.8cm}|p{3.2cm}|p{3.5cm}|p{3.5cm}|}
\hline
\textbf{Feature} & \textbf{Partial Homomorphic Encryption (PHE)} & \textbf{Somewhat Homomorphic Encryption (SWHE)} & \textbf{Fully Homomorphic Encryption (FHE)} \\ \hline
Supported Operations & Supports one type of operation (either addition or multiplication). & Supports both addition and multiplication but only for a limited number of operations (shallow circuit depth). & Supports arbitrary computations (both addition and multiplication without a fixed limit). \\ \hline
Noise Growth & Minimal noise accumulation, making these schemes computationally efficient. & Noise accumulates with each operation (linear for additions; quadratic or even exponential for multiplications), limiting computation depth. & Requires advanced noise management techniques (bootstrapping, modulus switching, relinearization) to control error growth over arbitrary computations. \\ \hline
Complexity \& Efficiency & Generally the most efficient and simplest in design (e.g., RSA, Paillier, ElGamal). & More complex than PHE; efficiency is affected by the need to manage noise, often limiting practical depth without additional techniques. & Most complex and computationally intensive due to elaborate noise control and the necessity for periodic refreshment of ciphertexts. \\ \hline
Common Examples & RSA (multiplicative), Paillier (additive), and ElGamal (multiplicative or adapted to additive in variants). & Early schemes such as Gentry’s original SWHE, later improved in the Brakerski--Gentry--Vaikuntanathan (BGV) and Fan--Vercauteren (FV) schemes. & Advanced lattice-based schemes including Gentry’s FHE, GSW, CKKS, among others, each incorporating techniques to enable fully homomorphic evaluation. \\ \hline
Security Assumptions & Relies on hardness problems such as the Integer Factorization Problem (IFP), Decisional Composite Residuosity (DCR), or the Discrete Logarithm Problem (DLP). & Based on lattice problems such as Learning With Errors (LWE) or Ring-LWE, with security intricately linked to noise management. & Built on advanced lattice assumptions (Ideal Lattice, LWE, RLWE, and variants) and often require additional measures to maintain security under extensive operations. \\ \hline
\end{tabular}

\label{tab:homomorphism}
\end{table}

\subsubsection{Partial Homomorphic Encryption(PHE)}
Partial Homomorphic Encryption (PHE) schemes support either additive or multiplicative homomorphism, allowing a single type of operation to be performed on ciphertexts. These schemes are computationally efficient but limited in functionality, as they cannot support both operations simultaneously. One of the most well-known PHE schemes is the RSA cryptosystem, which exhibits multiplicative homomorphism. In RSA, the product of two ciphertexts corresponds to the encryption of the product of their plaintexts, i.e., 
$Enc(m_1) \cdot Enc(m_2)=Enc(m_1\cdot m
_2)$. However, RSA in its textbook form is deterministic, making it semantically insecure. RSA's security relies on the Integer Factorization Problem (IFP), where the public key consists of a modulus $N=pq$, where $p,q$ are prime numbers, and an exponent $e$ coprime to $\phi(N)$. This simple version of RSA is vulnerable to chosen-ciphertext attacks if used without proper padding \cite{rivest1978data}.

Another widely used PHE scheme is the Paillier cryptosystem \cite{paillier1999public}, which supports additive homomorphism. In Paillier, the product of two ciphertexts corresponds to the encryption of the sum of their plaintexts, i.e., $Enc(m_1) + Enc(m_2)=Enc(m_1+ m
_2) \mod n^2$. Paillier is probabilistic, ensuring semantic security, and its hardness is based on the Decisional Composite Residuosity (DCR) assumption \cite{paillier1999public}. The public key in the Paillier cryptosystem consists of a modulus \( n = pq \). Encryption is performed by computing \( c = g^m \cdot r^n \mod n^2 \), where \( g \in \mathbb{Z}^*_{n^2} \) and \( 0 \le r \le n \) is a randomly chosen value such that \( gcd(r,n)=1 \). Paillier is particularly useful in applications such as electronic voting and privacy-preserving aggregation due to its additive properties.

The ElGamal cryptosystem and its variants, including EC-ElGamal, are also partially homomorphic encryption (PHE) schemes that exhibit multiplicative homomorphism. In ElGamal, the component-wise product of two ciphertexts corresponds to the encryption of the product of their plaintexts. Its security is based on the Discrete Logarithm Problem (DLP). The encryption process for plaintext $m$ involves computing the ciphertext \( c = (g^r, m \cdot h^r) \) using a randomly selected number $r$ where $1 \le r \le n$.  Here $g$ is the generator of a cyclic group $G$ of order $n$, $x$ is the private key and is a random element of the group and \( h = g^x \) is the public key \cite{elgamal1985public}. Variants like EC-ElGamal extend this framework to support additive homomorphism over elliptic curves, making them suitable for specific cryptographic applications.

Other noteworthy PHE schemes include the Modified RSA Encryption Algorithm (MREA) \cite{dhakar2012modified}, which extends RSA to support additive homomorphism, and the Chen-ElGamal (CEG) scheme \cite{hu2013improving}, which combines features of Paillier and ElGamal for hybrid operations. However, these schemes often face challenges related to inefficiency or ciphertext expansion, which can limit their practicality for large-scale applications.

\subsubsection{Somewhat Homomorphic Encryption (SWHE)}
Somewhat Homomorphic Encryption (SWHE) schemes enable additive and multiplicative operations on encrypted data; however, their computational utility is inherently limited by the growth of noise in the ciphertext. Each homomorphic operation introduces a small error term, and while additions increase noise linearly, multiplications typically cause quadratic (or even exponential) noise growth. Once the accumulated noise exceeds a certain threshold, correct decryption becomes impossible.

The concept of SWHE was first introduced by Gentry in 2009 with a scheme based on ideal lattices \cite{gentry2009fully}. In Gentry’s construction, the noise grows so rapidly with each multiplication that only circuits of very shallow depth can be evaluated reliably. To mitigate this, Gentry proposed the technique of bootstrapping — homomorphically evaluating the decryption function to refresh ciphertexts by reducing their noise. Despite its theoretical significance, the high computational cost of bootstrapping rendered early SWHE schemes impractical for many applications.

Subsequent research has focused on optimizing noise management to extend the viable depth of homomorphic computations without the need for frequent bootstrapping. The Brakerski–Gentry–Vaikuntanathan (BGV) scheme \cite{brakerski2012leveled} introduced modulus switching, a technique that reduces the ciphertext modulus and thereby decreases noise in a linear manner. This innovation enabled leveled homomorphic encryption, where the maximum allowable circuit depth is determined by the chosen parameters.

Similarly, the Fan–Vercauteren (FV) scheme \cite{fan2012somewhat}, which is based on the Ring Learning With Errors (RLWE) problem, employs efficient modulus reduction along with relinearization techniques to control noise growth. The FV scheme has proven particularly effective for applications such as encrypted database queries and privacy-preserving analytics \cite{boneh2013private}.

Further advancements were made by Brakerski’s 2012 work \cite{brakerski2012fully}, which introduced scale-invariant techniques. These schemes decouple noise accumulation from the magnitude of the plaintext, allowing the error to grow independently of the scale of the input data, thus enabling more efficient homomorphic computations.

More recently, alternative algebraic approaches have been explored to enhance both efficiency and security. For instance, a 2023 scheme based on random rank metric ideal codes \cite{aguilar2025somewhat} leverages the randomness inherent in code-based constructions to achieve competitive key and ciphertext sizes while supporting unlimited homomorphic additions and a fixed number of multiplications. In parallel, a scheme based on multivariate polynomial evaluation \cite{8733575} utilizes the algebraic properties of multivariate polynomials to perform limited-depth homomorphic operations with controlled noise growth. A recent scheme \cite{cryptoeprint:2024/1760} leverages the hardness of the sparse LPN (Learning Parities with Noise) problem combined with linear homomorphic properties. This construction supports the evaluation of bounded‐degree polynomials on encrypted data while maintaining compact ciphertext sizes. It represents a departure from traditional lattice‐based assumptions, thereby broadening the theoretical foundations for SWHE schemes.

Recent research has focused on refining noise management in SWHE schemes through optimized modulus switching and relinearization operations implemented in the Residue Number System (RNS) \cite{lee2023elasm}. By mapping computations into smaller rings and managing the scale more precisely, these techniques allow for tighter control over noise growth—thus supporting deeper circuit evaluations before bootstrapping becomes necessary. Such advancements have also influenced approximate schemes like CKKS \cite{cheon2017homomorphic}, which share similar noise management challenges to SWHE systems.

Although these advances have significantly improved the practicality of SWHE, noise management remains a central challenge. The trade-offs between computational depth, performance overhead, and security parameters continue to guide the design of modern SWHE systems, ensuring that even as new techniques are developed, careful attention must be paid to balancing noise accumulation against the desired homomorphic functionality.
\subsubsection{Fully Homomorphic Encryption (FHE)}
Gentry's model of homomorphic encryption relies on lattice-based cryptography. Over the past 15 years, there has been a significant surge in research and development focused on creating and enhancing homomorphic encryption schemes, resulting in a multitude of public key encryption methods capable of supporting homomorphic operations. Since the design of the first fully homomorphic encryption scheme, there have been nearly 40 public key schemes proposed, of  which only one is based on coding theory, 10 are based on number theory, and the rest are based on lattices. Current research focusses not only on the construction of homomorphic encryption but also on its application and implementation to various technologies.

A Fully Homomorphic Encryption (FHE) scheme is a type of encryption scheme that permits direct operations on ciphertexts, eliminating the need for decryption to perform operations. Homomorphic addition and multiplication operations are crucial to this functionality, since they constitute a functionally complete set within the realm of finite rings. Put differently, an FHE scheme enables the execution of computations on encrypted data without the need for prior decryption.
We can call any encryption scheme $E$ homomorphic with respect to any function $f$ if and only if the following statement is satisfied $$Dec(Eval(f,Enc(x),Enc(y)))= f(x,y), $$ where $x \text{ and } y $ are plain text.\\
Homomorphic encryption schemes possess two fundamental attributes: their ability to support a maximum degree of functions and the degree to which ciphertext length expands after each homomorphic operation. The first characteristic determines the types of functions the scheme can accurately evaluate. The second characteristic relates to the increase in ciphertext length, indicating how much the bit length of the ciphertext expands following each evaluation. When the limit on bit-length expansion remains consistent regardless of the complexity of the function, it is referred to as being compact. 

\subsection{The evolution of FHE schemes}
FHE schemes can be classified in 2017 into \textit{generations} based on the efficiency of computation and the techniques used to manage noise in evaluating the ciphertext \cite{cryptoeprint:2019/939}. This section describes the four generations of FHE schemes, the algorithms, their security assumptions, benefits and limitations. Table \ref{tab:FHE_models} summarises the comparison of the four generations of fully homomorphic encryption schemes. Note that code-based techniques have not been used in any of the four generations.


\begin{table}
\caption{Comparison of Fully Homomorphic Encryption (FHE) Models with respect to the underlying hard problem, the key techniques, advantages, disadvantages and whether bootstrapping is necessary.  }
\footnotesize
\begin{tabular}{|p{2cm}|p{2cm}|p{2cm}|p{2cm}|p{2cm}|p{2cm}|}
\hline
\textbf{FHE Model} & \textbf{Underlying Assumption / Hard Problem} & \textbf{Key Techniques} & \textbf{Advantages} & \textbf{Disadvantages} & \textbf{Bootstrapping Requirements} \\
\hline
First Generation: Ideal Lattice FHE (Gentry's Scheme) & Ideal lattice problems (e.g., SVP, SSSP, BDD) & Squashing to reduce decryption circuit complexity; Bootstrapping & Foundational breakthrough; supports arbitrary computations & High circuit complexity; challenging noise management; computationally expensive & Bootstrapping enabled via squashing (adds overhead) \\
\hline
First Generation: AGCD-Based FHE (DGHV Scheme) & Approximate GCD, combined with SSSP & Integer arithmetic with modular noise reduction & Conceptually simple; non-lattice alternative & Large public key size; high computational cost; less efficient in practice & Bootstrapping is required to control noise accumulation \\
\hline
Second Generation: LWE/RLWE-Based FHE (BV, BGV, FV Schemes) & Learning With Errors (LWE) and its ring variant (RLWE) & Modulus switching, relinearization, batching, scale invariance & More practical and efficient; implemented in libraries such as Microsoft SEAL & Noise growth limits circuit depth without bootstrapping; parameter tuning can be complex & Leveled FHE variants may avoid bootstrapping for fixed-depth circuits; bootstrapping still used for deeper evaluations \\
\hline
Second Generation: NTRU-Based FHE & NTRU assumption and circular security & Bootstrapping and modulus switching adapted to NTRU structure & Initially faster encryption operations & Vulnerabilities discovered; requires larger parameters for security; largely deprecated & Bootstrapping is incorporated, but overall robustness is lower \\
\hline
Third Generation: GSW-Based FHE (Approximate Eigenvector Method) & LWE/RLWE with an approximate eigenvector approach & Bit decomposition; optimized bootstrapping with reduced error growth & Improved noise management; supports deeper circuits with less noise amplification & Increased communication cost due to larger ciphertext sizes; higher computational overhead & Bootstrapping is more efficient, reducing overall complexity \\
\hline
Fourth Generation: CKKS Scheme & RLWE tailored for approximate arithmetic & Leveled encryption; plaintext embedding into complex number vectors; modulus scaling & Efficient for real-valued and approximate computations (e.g., ML applications) & Inherent approximation errors; requires careful precision management & Designed primarily as a leveled scheme (bootstrapping is optional) \\
\hline
\end{tabular}

\label{tab:FHE_models}
\end{table}

\subsubsection{First Generation: FHE based on Ideal Lattice}
The Gentry model \cite{gentry2009fully} represents the foundational framework for Fully Homomorphic Encryption (FHE), leveraging the hardness of problems in ideal lattices. The scheme operates over an ideal \( I \) and its corresponding integer sublattice \( L(I) \subseteq \mathbb{Z}^m \). 
The Gentry scheme is presented as below:
\begin{enumerate}
\item \textbf{Key Generation:}
\begin{itemize}
    \item Generate an ideal lattice $L$ with the secret basis $B_{sk}$ and a public basis $B_{pk}$.
    \item Select parameters: Modulus $q$, error distribution $\chi$.
\end{itemize}

    \item  \textbf{Encryption:}
    \begin{itemize}
    
   \item A message \( m \in \mathbb{F}_2 \) is encoded into a point \( a = m + 2e \in \mathbb{R}^d \), where \( e \) is a small error vector with coefficients sampled uniformly from \( \{0, \pm 1\} \). The probabilities of selecting \( +1 \) and \( -1 \) are equal.
   \item The ciphertext \( c \) is derived by translating \( a \) into the parallelepiped \( P(B_{pk}) \), where \( B_{pk} \) is the public key basis. This translation is achieved by computing:
     \[
     c = a - \lfloor a \cdot B_{pk}^{-1} \rceil \cdot B_{pk}.
     \]
   Where, \( \lfloor \cdot \rceil \) denotes rounding to the nearest integer.
\end{itemize}
\item \textbf{Decryption:}
\begin{itemize}
   \item To decrypt, the ciphertext \( c \) is first translated back into the parallelepiped \( P(B_{sk}) \), where \( B_{sk} \) is the secret key basis. This is done by computing:
     \[
     a' = c - \lfloor c \cdot B_{sk}^{-1} \rceil \cdot B_{sk}.
     \]
   \item The plaintext \( m \) is then recovered as \( m = a' \mod 2 \).
   \end{itemize}
\end{enumerate}
 The public key \( B_{pk} \) and secret key \( B_{sk} \) serve as bases for the ideal lattice \( L(I) \). The public key \( B_{pk} \) is constructed from skewed vectors, making it unsuitable as a basis for \( I \), whereas the secret key \( B_{sk} \) consists of orthogonal vectors, providing a suitable basis for \( I \).
 
Bootstrapping: The scheme is not inherently bootstrappable due to the high complexity of the decryption circuit. To address this, Gentry introduced the squashing technique, which reduces the decryption complexity by embedding additional information about the secret key into the evaluation key. 
 The security of the scheme relies on three core mathematical problems:

\begin{itemize}
  \item \textbf{Sparse Subset Sum Problem (SSSP):} Given a set of integers \( S = \{a_1, \dots, a_n\} \subseteq \mathbb{Z} \), determine whether there exists a sparse subset \( I \subseteq \{1, \dots, n\} \) such that \( \sum_{i \in I} a_i = 0 \).
  
  \item \textbf{Bounded Distance Decoding Problem (BDD):} Given a lattice and a target vector that is close to the lattice, find the closest lattice vector.
  
  \item \textbf{Ideal Shortest Vector Problem (Ideal SVP):} Find the shortest non-zero vector in an ideal lattice.
\end{itemize} 

Additionally, the scheme assumes circular security, which means that the encryption of the secret key does not compromise the overall security of the scheme.

\textbf{Limitations and Vulnerabilities:}
\begin{enumerate}
    \item \textbf{Efficiency Challenges:} Ideal lattice-based fully homomorphic encryption (FHE) schemes are computationally intensive due to the complex algebraic structures involved. Both the reliance on ideal lattices and the need for bootstrapping introduces significant overhead, limiting practical efficiency.
    
    \item \textbf{Principal Ideal Vulnerability:} Cramer et al. \cite{cramer2016recovering} identified a critical vulnerability in schemes based on principal ideal lattices. They demonstrated that key-recovery attacks are feasible if there exists a quantum polynomial-time or classical \( 2^{n^{\frac{2}{3} - \epsilon}} \)-time algorithm for solving the Principal Ideal Problem (PIP). This vulnerability undermines the security of schemes relying on principal ideal lattices, as an adversary could exploit the structure of the lattice to recover the secret key.
\end{enumerate}

\subsubsection{First Generation: FHE based on AGCD Problem}
In 2010, van Dijk et al. \cite{vandijk2010fully} introduced the DGHV Fully Homomorphic Encryption (FHE) scheme, which is built on number-theoretic assumptions. The DGHV scheme marks a significant milestone in the advancement of FHE, as it was the first construction to achieve fully homomorphic encryption using basic arithmetic operations over integers.

The DGHV scheme operates over the integers and relies on the hardness of the Approximate Greatest Common Divisor (AGCD) problem. The construction consists of three main algorithms: key generation, encryption, and decryption.

\begin{enumerate}
    \item \textbf{Key Generation:}
    \begin{itemize}
        \item The secret key is an odd random integer \( p \), sampled uniformly from a suitable range.
        \item The public key consists of a set of integers \( (x_0, x_1, \ldots, x_n) \), where each \( x_i \) is generated as:
        \[
        x_i = p \cdot q_i + r_i.
        \]
        Here, \( q_i \) is a large random integer, and \( r_i \) is a small random integer sampled from a noise distribution. The integer \( x_0 \) is chosen to be the largest odd integer in the public key.
    \end{itemize}
    \item \textbf{Encryption:}
    \begin{itemize}
        \item To encrypt a message \( m \in \mathbb{F}_2 \), the following steps are performed:
        \begin{enumerate}
            \item Select a random subset \( S \subseteq \{1, 2, \ldots, n\} \).
            \item Generate a random integer \( r \) from a noise distribution.
            \item Compute the ciphertext \( c \) as:
            \[
            c = m + 2r + 2 \sum_{i \in S} x_i \!\!\!\mod x_0.
            \]
        \end{enumerate}
        The ciphertext \( c \) is a noisy encoding of the message \( m \), where the noise terms \( r \) and \( r_i \) ensure security.
    \end{itemize}
    \item \textbf{Decryption:}
    \begin{itemize}
        \item To decrypt a ciphertext \( c \), the original message \( m \) is recovered as:
        \[
        m = (c\!\!\! \mod p)\!\!\! \mod 2.
        \]
        The decryption process relies on the fact that \( p \) is the secret key and the noise terms are small relative to \( p \).
    \end{itemize}
\end{enumerate} 

\textbf{Security Assumptions}

The security of the DGHV scheme is based on the following computational problems and assumptions:
\begin{itemize}
    \item \textbf{Approximate Greatest Common Divisor (AGCD) Problem:} Given a set of integers \( x_i = p \cdot q_i + r_i \), where \( p \) is a secret odd integer and \( r_i \) are small noise terms, the AGCD problem requires recovering \( p \). This problem is believed to be computationally hard, even for quantum computers.
    \item \textbf{Sparse Subset Sum Problem (SSSP):} The scheme also relies on the hardness of the SSSP, which involves finding a sparse subset of integers that sums to a specific value.
    \item \textbf{Circular Security:} The DGHV scheme assumes circular security, meaning that the encryption of the secret key \( p \) does not compromise the overall security of the scheme.
\end{itemize}

\textbf{Limitations and Drawbacks}

Despite its theoretical significance, the DGHV scheme has several practical limitations:
\begin{itemize}
    \item \textbf{Large Public Key Size:} The public key consists of \( n+1 \) integers, each of which is significantly larger than the secret key \( p \). This results in large public key size, which can be impractical for real-world applications.
    \item \textbf{High Computational Complexity:} The encryption and decryption processes involve arithmetic operations on large integers, leading to high computational overheads.
    \item \textbf{Noise Growth:} While the scheme supports homomorphic operations, the noise in the ciphertext grows with each operation, eventually requiring bootstrapping to maintain correctness. However, bootstrapping in the DGHV scheme is computationally expensive.
\end{itemize}
\subsubsection{Second Generation: FHE based on LWE and RLWE}
Brakerski and Vaikuntanathan introduced two Fully Homomorphic Encryption (FHE) schemes, marking the beginning of the second generation of FHE. These schemes are based on the Learning With Errors (LWE) problem \cite{brakerski2011efficient} and the Ring Learning With Errors (RLWE) problem \cite{brakerski2014leveled}. The symmetric scheme based on LWE, referred to as the BV scheme, and its RLWE-based counterpart, the BGV scheme, are foundational to modern FHE. Below, we provide a detailed description of these schemes, their algorithms, and their key innovations.\\

\noindent \textbf{BV Scheme: LWE-Based FHE}

The BV scheme is based on the LWE problem, which involves solving linear equations perturbed by small noise. The scheme consists of the following algorithms:

\begin{enumerate}
    \item \textbf{Encryption:}
    \begin{itemize}
        \item For a message \( m \in \mathbb{F}_2 \), the ciphertext \( c \) is represented as:
        \[
        c = (a, b) = \left(a, \langle a, s \rangle + 2e + m\right),
        \]
        where:
        \begin{itemize}
            \item \( a \) is a random vector in \( \mathbb{Z}_q^n \),
            \item \( s \) is the secret key, a random vector in \( \mathbb{Z}_q^n \),
            \item \( e \) is a small error term drawn from an error distribution \( \chi \),
            \item \( \langle a, s \rangle \) denotes the inner product of \( a \) and \( s \).
        \end{itemize}
    \end{itemize}

    \item \textbf{Decryption:}
    \begin{itemize}
        \item The decryption process computes the plaintext as:
        \[
        m = \left(b - \langle a, s \rangle \!\!\! \mod q\right)\!\!\! \mod 2.
        \]
        Decryption succeeds if the error term \( e \) is sufficiently small, specifically when \( |e| < q/4 \).
    \end{itemize}
\end{enumerate}

The BV scheme introduced two critical techniques to enable fully homomorphic encryption:
\begin{itemize}
    \item \textbf{Relinearization (Key-Switching):} This technique reduces the size of ciphertexts after homomorphic multiplication. Specifically, it reduces the ciphertext size from \( \mathcal{O}(n^2) \) to \( \mathcal{O}(n) \), where \( n \) is the dimension of the LWE problem.
    \item \textbf{Dimension-Modulus Reduction (Modulus Switching):} This technique transforms a ciphertext \( c'\!\!\! \mod q \) into a ciphertext \( c \!\!\! \mod p \), where \( p \ll q \). Modulus switching reduces the noise growth during homomorphic operations, enabling deeper computations without bootstrapping.
\end{itemize}

By eliminating the need for Gentry's squashing technique as well as the reliance on the Sparse Subset Sum Problem (SSSP), the BV scheme becomes more efficient and practical.\\

\noindent \textbf{BGV Scheme: RLWE-Based FHE}

The BGV scheme extends the BV framework to the Ring Learning With Errors (RLWE) setting, offering improved efficiency and scalability. The scheme operates over the ring \( R = \mathbb{Z}[x]/\langle x^d + 1 \rangle \), where \( d = 2^M \) for some integer \( M \). The BGV scheme is defined as follows:

\begin{enumerate}
    \item \textbf{Key Generation:}
    \begin{itemize}
        \item A secret key \( sk = (1, s) \in R_q^2 \) is generated, where \( s \) is a small random element sampled from the error distribution \( \chi \).
        \item A public key \( pk = (a, b) \) is computed as:
        \[
        b = a \cdot s + 2e \!\!\!\mod q,
        \]
        where \( a \) is a random element in \( R_q \), and \( e \) is a small error term.
    \end{itemize}

    \item \textbf{Encryption:}
    \begin{itemize}
        \item For a message \( \mu \in \mathbb{F}_2 \), the message is encoded as \( m = (\mu, 0) \in R_q^2 \).
        \item The ciphertext \( c \) is computed as:
        \[
        c = m + 2(e_0, e_1) + a \cdot r,
        \]
        where \( r, e_0, e_1 \) are small random elements sampled from \( \chi \).
    \end{itemize}

    \item \textbf{Decryption:}
    \begin{itemize}
        \item The decryption process computes the inner product \( \langle c, s \rangle \) and outputs the plaintext as:
        \[
        m = \langle c, s \rangle \!\!\! \mod 2.
        \]
    \end{itemize}
\end{enumerate}

Brakerski introduced a scale-invariant variant of the BGV scheme \cite{brakerski2012fully}, achieved by scaling down both the ciphertext and the error by a factor of \( q \), where \( q \) is the ciphertext modulus, thus reducing the noise growth during homomorphic multiplications from exponential to linear.  This innovation replaces the traditional modulus switching technique, further improving the scheme's efficiency.\\

\noindent \textbf{FV Scheme: Optimized RLWE-Based FHE}

The FV scheme, an optimized version of the BGV scheme, was introduced by Fan and Vercauteren \cite{fan2012somewhat}. It is designed for efficient modular arithmetic on encrypted integers and is implemented in Microsoft's SEAL library \cite{sealcrypto}. The FV scheme operates as follows:

\begin{enumerate}
    \item \textbf{Key Generation:}
    \begin{itemize}
        \item The secret key \( sk = s \) is sampled from the error distribution \( \chi \).
        \item The public key \( pk = (p_0, p_1) \) is computed as:
        \[
        p_0 = -(a \cdot s + e) \!\!\! \mod q, \quad p_1 = a,
        \]
        where \( a \) is a random element in \( R_q \), and \( e \) is a small error term.
    \end{itemize}

    \item \textbf{Encryption:}
    \begin{itemize}
        \item For a message \( m \in R_p \), the ciphertext \( c = (c_0, c_1) \) is computed as:
        \[
        c_0 = (p_0 \cdot u + e_1 + \Delta m) \!\!\! \mod q, \quad c_1 = (p_1 \cdot u + e_2)\!\!\!  \mod q,
        \]
        where \( u, e_1, e_2 \) are small random elements sampled from \( \chi \), and \( \Delta = \lfloor q/p \rfloor \).
    \end{itemize}

    \item \textbf{Decryption:}
    \begin{itemize}
        \item The decryption process computes:
        \[
        m = \left\lfloor \frac{p \cdot (c_0 + c_1 \cdot s) \mod q}{q} \right\rfloor \!\!\!\mod p.
        \]
    \end{itemize}
\end{enumerate}

The BGV and FV schemes have been further enhanced through techniques such as batching, modulus switching, and improved bootstrapping. These optimizations enable their integration into practical cryptographic libraries like HElib \cite{cryptoeprint:2020/1481} and SEAL \cite{sealcrypto}, making them suitable for applications such as privacy-preserving machine learning and secure computation.

\subsubsection{Second Generation: FHE based on NTRU}
The NTRU encryption scheme, introduced by Hoffstein et al. in 1998 \cite{hoffstein1998ntru}, is a lattice-based cryptosystem that has played a significant role in the development of post-quantum cryptography. The scheme was initially proposed with a provisional patent, which was later granted in 2000 \cite{hoffstein2000public}. NTRU is often referred to as the LTV scheme in the literature \cite{lopez2012on}, and it incorporates techniques such as bootstrapping and modulus switching to achieve its functionality.

The NTRU scheme operates over the ring \( R = \mathbb{Z}[x]/\langle x^d + 1 \rangle \), where \( d = 2^m \) for some integer \( m \). The scheme consists of three main algorithms: key generation, encryption, and decryption.

\begin{enumerate}
    \item \textbf{Key Generation:}
    \begin{itemize}
        \item Let \( \chi \) be a bounded distribution over the ring \( R \). Sample two small random polynomials \( f' \) and \( g \) from \( \chi \).
        \item Compute the secret key polynomial \( f = 2f' + 1 \), ensuring that \( f \equiv 1 \!\! \mod 2 \) and is invertible in \( R_q = \mathbb{Z}_q[x]/\langle x^d + 1 \rangle \).
        \item Compute the public key polynomial \( h = 2g \cdot f^{-1} \!\!\!\mod q \).
        \item The key pair is \( (pk, sk) = (h, f) \).
    \end{itemize}

    \item \textbf{Encryption:}
    \begin{itemize}
        \item For a binary message \( m \in \{0, 1\} \), sample small random polynomials \( s \) and \( e \) from \( \chi \).
        \item Compute the ciphertext \( c \) as:
        \[
        c = h \cdot s + 2e + m \!\!\!\mod q.
        \]
        Here, \( h \) is the public key, and \( s \) and \( e \) are small random noise terms.
    \end{itemize}

    \item \textbf{Decryption:}
    \begin{itemize}
        \item To decrypt the ciphertext \( c \), compute:
        \[
        m = (f \cdot c \!\!\!\mod q) \!\!\!\mod 2.
        \]
        The correctness of decryption relies on the fact that \( f \) is the secret key and that the noise terms \( s \) and \( e \) are small.
    \end{itemize}
\end{enumerate}

\noindent \textbf{Security Assumptions}

The security of the NTRU scheme is based on the following computational problems and assumptions:
\begin{itemize}
    \item \textbf{Circular Security:} The scheme assumes that the encryption of the secret key does not compromise the overall security of the system.
    \item \textbf{Ring Learning With Errors (RLWE) Problem:} The hardness of solving the RLWE problem over the ring \( R \) is a fundamental assumption for the security of NTRU.
    \item \textbf{Decisional Small Polynomial Ratio (DSPR) Problem:} This problem involves distinguishing between a random polynomial and a ratio of two small polynomials in the ring \( R \). The DSPR problem is believed to be computationally hard, even for quantum computers.
\end{itemize}

\noindent \textbf{Limitations and Vulnerabilities}

Despite its initial promise, the NTRU scheme has several limitations and vulnerabilities that have impacted its practical adoption:
\begin{itemize}
    \item \textbf{Parameter Sizing:} To achieve security against known attacks, the parameters of NTRU-based schemes must be significantly larger than those initially proposed. This results in larger key sizes and reduced efficiency.
    \item \textbf{Error Rates:} Research by Lepoint and Naehrig \cite{LN14} has shown that RLWE-based schemes exhibit lower error rates compared to NTRU-based schemes. This makes RLWE-based schemes more robust for practical applications.
    \item \textbf{Attacks:} Numerous attacks have been documented against NTRU-based schemes, including lattice reduction attacks and subfield attacks. While parameters have been improved to mitigate these attacks, the resulting schemes are less efficient than their RLWE counterparts.
\end{itemize}

Due to the aforementioned limitations, NTRU-based schemes are no longer widely used or supported by cryptographic libraries. The efficiency and security advantages of RLWE-based schemes have made them the preferred choice for lattice-based cryptography in both academic research and practical implementations.

\subsubsection{Third Generation: FHE Based on LWE and RLWE}
The third generation of homomorphic encryption introduced a novel approach known as the approximate eigenvector method \cite{gentry2013homomorphic}. This method significantly advances the field by eliminating the need for key and modulus switching techniques, which were critical in earlier schemes such as BGV and FV. A key advantage of this approach is its ability to control error growth during homomorphic multiplications, limiting it to a small polynomial factor. Specifically, when multiplying \( l \) ciphertexts with the same error level, the final error increases by a factor of \( l \cdot \text{poly}(n) \), where \( n \) represents the dimension of the encryption scheme or lattice. This is a substantial improvement over previous schemes, where error growth was quasi-polynomial. This third-generation scheme is commonly referred to as the GSW scheme, named after its inventors Gentry, Sahai, and Waters. Below, we provide a detailed description of the GSW scheme, its algorithms, and its optimizations.\\

\noindent \textbf{GSW Scheme: Approximate Eigenvector Method}

The GSW scheme is based on the Learning With Errors (LWE) problem and employs an approximate eigenvector approach to achieve fully homomorphic encryption. The scheme consists of the following algorithms:

\begin{enumerate}
    \item \textbf{Key Generation:}
    \begin{itemize}
        \item The secret key \( s \) is a vector of the form \( s = (1, s_2, \dots, s_n) \in \mathbb{Z}_q^n \), where each \( s_i \) is randomly selected.
        \item The public key \( A \in \mathbb{Z}_q^{n \times n} \) is constructed such that \( A \cdot s \approx 0 \). This ensures that the product of \( A \) and the secret key \( s \) results in a small error term.
    \end{itemize}

    \item \textbf{Encryption:}
    \begin{itemize}
        \item For a message \( m \in \mathbb{Z}_q \), the ciphertext \( C \) is constructed as:
        \[
        C = m I_n + R A,
        \]
        where:
        \begin{itemize}
            \item \( I_n \) is the \( n \times n \) identity matrix,
            \item \( R \) is a random binary matrix of dimension \( n \times n \),
            \item \( A \) is the public key.
        \end{itemize}
        The ciphertext \( C \) is a noisy encoding of the message \( m \), where the noise is introduced by the matrix \( R \).
    \end{itemize}

    \item \textbf{Decryption:}
    \begin{itemize}
        \item To decrypt the ciphertext \( C \), compute:
        \[
        C \cdot s = m I_n s + R A s \approx m I_n s + R e \approx m I_n s,
        \]
        where \( e \) is a small error term. Since \( s_1 = 1 \), the first element of the vector \( C \cdot s \) approximates \( m \).
        \item The decrypted message is obtained as:
        \[
        m \approx (C \cdot s)_1.
        \]
    \end{itemize}
\end{enumerate}

\noindent \textbf{Homomorphic Multiplication and Bootstrapping}

The GSW scheme employs bit decomposition for homomorphic multiplication allowing for efficient computation of products of ciphertexts. However, the scheme has some limitations:
\begin{itemize}
    \item \textbf{High Communication Costs:} The ciphertext size in the GSW scheme is relatively large compared to the plaintext, leading to increased communication overhead.
    \item \textbf{Computational Complexity:} The scheme involves complex matrix operations, which can be computationally intensive.
\end{itemize}

To address these limitations, several optimizations have been proposed:
\begin{itemize}
    \item \textbf{Arithmetic Bootstrapping:} Alperin-Sheriff and Peikert (AP) \cite{alperin2014faster} introduced a bootstrapping algorithm that treats decryption as an arithmetic function rather than a Boolean circuit, significantly reducing computational complexity .
    \item \textbf{Homomorphic Matrix Operations:} Hiromasa et al. \cite{Hiromasa2015} extended the GSW scheme to support homomorphic matrix operations, further enhancing its applicability.
    \item \textbf{Programmable Bootstrapping (PBS):} Ducas and Micciancio \cite{ducas2015fhew} introduced the FHEW scheme, which incorporates the AP bootstrapping technique and enables programmable bootstrapping. This allows for the homomorphic evaluation of arbitrary functions, including the NAND operation, using lookup tables.
\end{itemize}

In addition to the GSW scheme, the third generation of homomorphic encryption includes three distinct schemes based on the torus:
\begin{itemize}
    \item \textbf{TLWE (Torus Learning With Errors):} Generalizes the LWE problem to the torus \( \mathbb{T} = \mathbb{R}/\mathbb{Z} \).
    \item \textbf{TRLWE (Torus Ring Learning With Errors):} Extends the RLWE problem to the torus, enabling encryption of plaintexts in the ring of integer polynomials \( R \).
    \item \textbf{TRGSW (Torus Ring GSW):} A ring variant of the GSW scheme on the torus, which supports homomorphic operations on encrypted data.
\end{itemize}

The TRLWE scheme operates over the ring \( R = \mathbb{Z}[x]/\langle x^d + 1 \rangle \), where \( d = 2^M \). The message space is defined as \( T = R[x]/\langle x^d + 1 \rangle \mod 1 \), and the scheme is constructed as follows:

\begin{enumerate}
    \item \textbf{Key Generation:}
    \begin{itemize}
        \item For a security parameter \( \lambda \), generate a small secret key \( s \in R_2^n \), where \( R_2 = \mathbb{F}_2[x]/\langle x^d + 1 \rangle \).
    \end{itemize}

    \item \textbf{Encryption:}
    \begin{itemize}
        \item For a message \( m \in \mathbb{T} \), sample a random mask \( a \) uniformly from \( \mathbb{T}^n \) and a small error \( e \) from a bounded distribution \( \chi \).
        \item Compute the ciphertext \( c \) as:
        \[
        c = (a, s \cdot a + m + e) \in \mathbb{T}^n \times \mathbb{T}.
        \]
    \end{itemize}

    \item \textbf{Decryption:}
    \begin{itemize}
        \item Compute the phase function \( \psi_s(c) = b - s \cdot a \), where \( c = (a, b) \).
        \item The decrypted message is obtained by rounding \( \psi_s(c) \) to the nearest point in the message space \( M \subseteq \mathbb{T} \).
    \end{itemize}
\end{enumerate}

Chillotti et al. \cite{chillotti2021improved} introduced several optimizations for torus-based FHE schemes, including the ability to generate fresh ciphertexts by linearly combining existing ones. However, non-linear operations remain challenging due to the limitations of TLWE. To address this, the TRGSW scheme was developed, which supports generalized scale-invariant operations and enables efficient switching between TRLWE and TLWE.

\subsubsection{Fourth Generation: FHE based on LWE and RLWE} In 2017, Cheon et al. introduced a groundbreaking collection of Fully Homomorphic Encryption (FHE) schemes \cite{cheon2017homomorphic}, which are specifically designed for approximate arithmetic on real and complex numbers. This scheme, originally named HEAAN (Homomorphic Encryption for Arithmetic of Approximate Numbers), is now commonly referred to as the CKKS scheme, named after its authors. The CKKS scheme provides a leveled homomorphic encryption framework, enabling efficient computations on encrypted data while maintaining approximate correctness. Additionally, the authors released an open-source library to facilitate the implementation and deployment of this scheme.

The CKKS scheme operates over the ring \( R = \mathbb{Z}[x]/\langle x^d + 1 \rangle \), where \( d = 2^M \) for some integer \( M \). The scheme is designed to handle approximate arithmetic, making it particularly suitable for applications involving real or complex numbers, such as machine learning and scientific computing. The CKKS scheme consists of the following algorithms:

\begin{enumerate}
    \item \textbf{Key Generation:}
    \begin{itemize}
        \item Let \( p \) be a base, \( q_0 \) a modulus, and \( L \) a selected level. Define \( q_l = p_l \cdot q_0 \) for \( l = 1, \dots, L \).
        \item Let \( \chi \) be a \( B \)-bounded distribution, and let \( DG(\sigma^2) \) denote a vectorial discrete Gaussian distribution over \( \mathbb{Z}^d \) with variance \( \sigma^2 \). Let \( ZO(\rho) \) denote a distribution over \( \{-1, 0, 1\}^d \) with \( 0 < \rho < 1 \).
        \item Generate the secret key \( sk = (1, s) \), where \( s \) is sampled from \( \chi \).
        \item Sample a random value \( a \in R_{q_L} \) and compute \( b = -a \cdot s + e \mod q_L \), where \( e \) is sampled from \( DG(\sigma^2) \).
        \item Sample \( a' \in R_{t \cdot q_L} \) and \( e' \in DG(\sigma^2) \); compute \( b' = -a \cdot s + e' + t \cdot s'\!\!\!  \mod t \cdot q_L \).
        \item Output the secret key \( sk = (1, s) \), the public key \( pk = (b, a) \), and the evaluation key \( evk = (b', a') \).
    \end{itemize}

    \item \textbf{Encryption:}
    \begin{itemize}
        \item For a message \( m \in R \), sample random values \( v \in ZO(1/2) \) and \( e_0, e_1 \in DG(\sigma^2) \).
        \item Compute the ciphertext \( c = (\beta, \alpha) \), where:
        \[
        \beta = v \cdot b + m + e_0 \!\!\! \mod q_l, \quad \alpha = v \cdot a + e_1 \!\!\! \mod q_l.
        \]
        The ciphertext \( c \) is a noisy encoding of the message \( m \), where the noise terms \( e_0 \) and \( e_1 \) ensure security.
    \end{itemize}

    \item \textbf{Decryption:}
    \begin{itemize}
        \item To decrypt the ciphertext \( c = (\beta, \alpha) \), compute:
        \[
        m = \langle c, sk \rangle \!\!\! \mod q_l = \beta + \alpha \cdot s \!\!\! \mod q_l.
        \]
        The decrypted message \( m \) is an approximation of the original plaintext, with the error introduced during encryption.
    \end{itemize}
\end{enumerate}

A notable feature of the CKKS scheme is its ability to encode messages as elements in the extension field \( \mathbb{C} \), which corresponds to the complex numbers. Specifically:
\begin{itemize}
    \item The message space is defined as \( S = R[x]/\langle x^d + 1 \rangle \), where the roots of the polynomial \( x^d + 1 \) are the complex primitive roots of unity in \( \mathbb{C} \).
    \item A message \( m \in S \) can be embedded into a vector of complex numbers by evaluating it at these roots. This allows for efficient encoding and decoding of real and complex numbers, making the CKKS scheme particularly suitable for applications involving approximate arithmetic.
\end{itemize}

The CKKS scheme offers several advantages over previous FHE schemes:
\begin{itemize}
    \item \textbf{Approximate Arithmetic:} The scheme is specifically designed for approximate computations, making it ideal for applications such as machine learning, where exact precision is not required.
    \item \textbf{Efficient Encoding:} The ability to encode messages as complex numbers allows for efficient representation of real and complex data, reducing the computational overhead associated with homomorphic operations.
    \item \textbf{Leveled Homomorphism:} The CKKS scheme supports leveled homomorphic encryption, enabling a predetermined number of homomorphic operations without the need for bootstrapping. This makes it more efficient for practical applications.
\end{itemize}

\subsection{Security Assumptions of current FHE schemes}

The security of lattice-based Fully Homomorphic Encryption (FHE) schemes relies on the hardness of several well-studied computational problems in lattice theory. These problems form the foundation of the cryptographic assumptions underlying the security of LWE (Learning With Errors) and RLWE (Ring Learning With Errors)-based schemes. Below, we provide a brief analysis of the various problems and their relevance to FHE. The reader can find more details about the NP-Hard problems in \cite{li2022tutorial}.

\subsubsection{Shortest Vector Problem (SVP)}

The Shortest Vector Problem (SVP) is a fundamental problem in lattice theory and serves as the basis for many lattice-based cryptographic schemes. Given a lattice \( L \), the SVP involves finding the shortest non-zero vector in \( L \). Several variants of the SVP are used in cryptographic constructions:

\begin{itemize}
    \item \textbf{\(\gamma\)-Approximate Shortest Vector Problem (\(SVP_{\gamma}\)):}
    \begin{itemize}
        \item Given a lattice \( L \) and an approximation factor \( \gamma \geq 1 \), the goal is to find a non-zero vector \( v \in L \) such that:
        \[
        \|v\| \leq \gamma \cdot \lambda_1(L),
        \]
        where \( \lambda_1(L) \) is the length of the shortest non-zero vector in \( L \).
    \end{itemize}

    \item \textbf{Decisional Shortest Vector Problem (\(GapSVP_{\gamma,r}\)):}
    \begin{itemize}
        \item Given a lattice \( L \), an approximation factor \( \gamma \geq 1 \), and a bound \( r > 0 \), the goal is to decide whether:
        \[
        \lambda_1(L) \leq r \quad \text{or} \quad \lambda_1(L) \geq \gamma \cdot r.
        \]
    \end{itemize}

    \item \textbf{\(\gamma\)-Unique Shortest Vector Problem (\(uSVP_{\gamma}\)):}
    \begin{itemize}
        \item Given a lattice \( L \) and an approximation factor \( \gamma \geq 1 \), the goal is to find the shortest non-zero vector \( v \in L \) under the condition that:
        \[
        \lambda_1(L) < \gamma \cdot \lambda_2(L),
        \]
        where \( \lambda_2(L) \) is the length of the second shortest vector in \( L \).
    \end{itemize}
\end{itemize}

The hardness of these problems is crucial for the security of lattice-based FHE schemes. If the SVP is solvable, then the decisional variant \( GapSVP_{\gamma,r} \) is also solvable.

\subsubsection{Closest Vector Problem (CVP) }

The Closest Vector Problem (CVP) is a generalization of the SVP, where the goal is to find the lattice vector closest to a given target vector \( t \in \mathbb{R}^n \). The CVP and its variants are defined as follows:

\begin{itemize}
    \item \textbf{\(\gamma\)-Approximate Closest Vector Problem (\(CVP_{\gamma}\)):}
    \begin{itemize}
        \item Given a lattice \( L \), a target vector \( t \in \mathbb{R}^n \), and an approximation factor \( \gamma \geq 1 \), the goal is to find a vector \( v \in L \) such that:
        \[
        \text{dist}(t, v) \leq \gamma \cdot \text{dist}(t, L),
        \]
        where \( \text{dist}(t, L) \) is the distance between \( t \) and the lattice \( L \).
    \end{itemize}

    \item \textbf{Decisional Closest Vector Problem (\(DCVP_{\gamma,r}\)):}
    \begin{itemize}
        \item Given a lattice \( L \), a target vector \( t \in \mathbb{R}^n \), an approximation factor \( \gamma \geq 1 \), and a bound \( r > 0 \), the goal is to decide whether:
        \[
        \text{dist}(t, L) \leq r \quad \text{or} \quad \text{dist}(t, L) \geq \gamma \cdot r.
        \]
    \end{itemize}

    \item \textbf{\(\alpha\)-Bounded Distance Decoding (\(BDD_{\alpha}\)):}
    \begin{itemize}
        \item Given a lattice \( L \), a target vector \( t \in \mathbb{R}^n \), and a parameter \( \alpha \leq 1 \), the goal is to find the closest lattice vector \( v \in L \) under the condition that:
        \[
        \text{dist}(t, L) < \alpha \cdot \lambda_1(L).
        \]
    \end{itemize}
\end{itemize}

The CVP and its variants are closely related to the SVP and are used in the security analysis of lattice-based cryptographic schemes.

\subsubsection{Shortest Integer Solution (SIS) Problem}

The Shortest Integer Solution (SIS) problem is another fundamental problem in lattice-based cryptography. It is defined as follows:

\begin{itemize}
    \item \textbf{SIS Problem (\(SIS_{q,m,\beta}\)):}
    \begin{itemize}
        \item Given a matrix \( A \in \mathbb{Z}_q^{m \times n} \) and a bound \( \beta < q \), the goal is to find a non-zero integer vector \( x \in \mathbb{Z}^m \) such that:
        \[
        xA \equiv 0 \pmod{q} \quad \text{and} \quad \|x\| \leq \beta.
        \]
    \end{itemize}
\end{itemize}

The SIS problem is equivalent to finding a short vector in the scaled dual lattice \( L_q^\perp(A) \). It is used in the construction of cryptographic primitives such as digital signatures and hash functions.

\subsubsection{Learning With Errors (LWE) and Ring-LWE}

The Learning With Errors (LWE) problem and its ring variant (Ring-LWE) are central to the security of many lattice-based FHE schemes. These problems are defined as follows:

\begin{itemize}
    \item \textbf{LWE Problem:}
    \begin{itemize}
        \item Given a matrix \( A \in \mathbb{Z}_q^{m \times n} \), a secret vector \( s \in \mathbb{Z}_q^n \), and a noise vector \( e \in \mathbb{Z}_q^m \) sampled from a discrete Gaussian distribution, the LWE instance is:
        \[
        b = As + e \pmod{q}.
        \]
        The goal is to recover the secret vector \( s \) given \( (A, b) \).
    \end{itemize}

    \item \textbf{Ring-LWE Problem:}
    \begin{itemize}
        \item Given a polynomial ring \( R_q = \mathbb{Z}[x]/\langle f(x) \rangle \), a secret polynomial \( s(x) \in R_q \), and a noise polynomial \( e(x) \in R_q \), the Ring-LWE instance is:
        \[
        b(x) = a(x)s(x) + e(x) \pmod{f(x)},
        \]
        where \( a(x) \in R_q \) is a random polynomial. The goal is to recover the secret polynomial \( s(x) \) given \( (a(x), b(x)) \).
    \end{itemize}
\end{itemize}

The security of LWE and Ring-LWE-based schemes relies on the hardness of solving these problems using lattice reduction algorithms. However, as demonstrated by Albrecht et al. \cite{castryck2016error}, there is no universal attack that can efficiently solve all instances of these problems. The effectiveness of lattice reduction attacks depends on the choice of parameters, such as the modulus \( q \), the dimension \( n \), and the error distribution.

For Ring-LWE, the security analysis is similar to that of LWE, but with additional considerations due to the ring structure. According to the Homomorphic Encryption Security Standard, selecting an appropriate error distribution ensures that there are no superior attacks on Ring-LWE compared to LWE. However, the error distribution must be sufficiently spread to maintain security.

The security of lattice-based FHE schemes is grounded in the hardness of problems such as SVP, CVP, SIS, LWE, and Ring-LWE. These problems are believed to be resistant to quantum attacks, making lattice-based cryptography a promising candidate for post-quantum security. While lattice reduction algorithms pose a potential threat, careful parameter selection and error distribution management can mitigate these risks, ensuring the robustness of lattice-based cryptographic schemes.

Lattice-based homomorphic encryption (HE) schemes, like those based on Learning With Errors (RLWE) and NTRU, have made notable strides in efficiency and security. However, they still face challenges such as large ciphertext expansion, high computational overhead, and susceptibility to quantum attacks. Concerns about their long-term security arise from their reliance on lattice problems.
To address these limitations, researchers are turning to alternative hardness assumptions. Code-based cryptography, known for its strength in post-quantum encryption, offers potential benefits for HE, including robust security and efficiency. Exploring code-based HE schemes may help overcome the shortcomings of lattice approaches while maintaining resistance to quantum threats.

\section{Code-Based Cryptography}\label{Sec:CBC}

Code-based cryptography relies on the NP-hard problem of indistigushabilty and decoding a random linear code and is a promising option for post-quantum cryptography. Code-based cryptosystems work by choosing the private key as a linear code $C$, which can efficiently correct a predetermined number of errors. The public key, denoted as $C'$, is a disguised version of the linear code and is designed so that it will appear random and not reveal the private key to an observer who doesn't have access to $C$. With access to the public key $C'$, a sender can encrypt their message and potentially add errors. An attacker who sees the ciphertext, will need to decode it to recover the message. As the ciphertext is a random codeword, there is no polynomial-time algorithm to do this. The receiver who has the private key $C$ can transform the ciphertext into a codeword of $C$, which can be efficiently decoded to retrieve the message.\\
In this section we first introduce the NP-hard problems on which code-based cryptography is based. We will then introduce existing code-based ciphers.

\subsection{NP-Hard Problems in Coding Theory}
There are two categories of NP-hard problems in this area. The first is around the difficulty of decoding a random linear code and the second is around the difficulty of proving equivalence of codes. While the NP-hard problems in both categories are defined over the Hamming metric, we also present the analogous problems in the rank metric.

\subsubsection{ The Decoding Problems in Coding Theory}
There are three variations of decoding problems, each of which is described below:\\

\noindent \textbf{Decoding Problem (DP):}\\
Let $\mathbb{F}_q$ be a finite field and consider a code defined by a generator matrix 
\[
G \in \mathbb{F}_q^{k \times n}.
\]
Given a received vector $r \in \mathbb{F}_q^n$ and an integer $t$ (representing an error weight threshold), the task is to decide whether there exists a message $m \in \mathbb{F}_q^k$ and an error vector $e \in \mathbb{F}_q^n$ with 
\[
\operatorname{wt}(e) \le t,\;
\text{such that,} \;\;
r = mG + e.
\]
This problem is fundamental in algebraic coding theory.

\vspace{0.5em}
\noindent \textbf{Syndrome Decoding Problem (SDP):}\\
In the syndrome formulation, one is given a parity-check matrix 
\[
H \in \mathbb{F}_q^{(n-k) \times n},
\]
a syndrome $s \in \mathbb{F}_q^{n-k}$, and an integer $t$. The goal is to find an error vector $e \in \mathbb{F}_q^n$ satisfying
\[
e H^\top = s \quad \text{and} \quad \operatorname{wt}(e) \le t.
\]
By converting $G$ into systematic form and obtaining the corresponding $H$, the DP can be recast as an SDP, and conversely, one can recover a DP instance from an SDP instance.

\vspace{0.5em}
\noindent \textbf{Given Weight Codeword Problem (GWCP):} \\
Given a parity-check matrix $H$ and an integer $w$, the problem asks whether there exists a codeword 
\[
c \in \mathbb{F}_q^n,
\]
such that 
\[
cH^\top = 0 \quad \text{and} \quad \operatorname{wt}(c) = w.
\]
By augmenting the generator matrix with the received vector, one shows that GWCP is equivalent to the DP (and hence to the SDP).

\subsubsection{Code Equivalence Problems}
Code equivalence problem is known as the indistinguishability of a random linear code.
Similar to decoding problems, there are three varieties of code equivalence problems.  \\

\noindent \textbf{Permutation Equivalence Problem (PEP):}\\
Given two generator matrices 
\[
G,\, G' \in \mathbb{F}_q^{k \times n},
\]
find a permutation $\phi \in S_n$ (the symmetric group on $n$ elements) such that
\[
\phi(\langle G \rangle) = \langle G' \rangle.
\]
This problem is a special case of the broader linear equivalence issues.

\vspace{0.5em}
\noindent \textbf{Linear Equivalence Problem (LEP):}\\
For $G,\, G' \in \mathbb{F}_q^{k \times n}$, the goal is to find a mapping 
\[
\phi \in (\mathbb{F}_q^\star)^n \rtimes S_n,
\]
that sends the code generated by $G$ to that generated by $G'$.

\vspace{0.5em}
\noindent \textbf{Subcode Equivalence and the Permuted Kernel Problem (PKP/SEP):}
\begin{itemize}
    \item \textbf{PKP:} Given $G \in \mathbb{F}_q^{k \times n}$ and another matrix $H'$ (typically related to a subcode), find a permutation matrix $P$ such that
    \[
    H'(GP)^\top = 0.
    \]
    \item \textbf{SEP:} Reformulated as the subcode equivalence problem, one seeks a permutation matrix $P$ such that
    \[
    \langle G' \rangle \subset \langle GP \rangle.
    \]
\end{itemize}
A relaxed version of PKP requires only finding a non-zero codeword (i.e. a subcode of dimension 1) that meets the equivalence condition.

\subsubsection{ Rank-Metric Analogues and the MinRank Problem}
There are analogues to the problems above with respect to the rank metric. \\
\noindent \textbf{Rank Syndrome Decoding Problem (Rank SDP):}\\
In the rank metric, where matrices are considered over $\mathbb{F}_{q^m}$, the problem is analogous to the classical SDP. Given a parity-check matrix 
\[
H \in \mathbb{F}_{q^m}^{(n-k) \times n},
\]
a syndrome 
\[
s \in \mathbb{F}_{q^m}^{n-k},
\]
and an integer $t$, the task is to find an error vector 
\[
e \in \mathbb{F}_{q^m}^n,
\]
with rank weight 
\[
\operatorname{wt}_R(e) \le t,
\]
satisfying 
\[
e H^\top = s.
\]
While reductions from the classical SDP suggest its hardness, the NP-completeness of the Rank SDP remains an open question.

\vspace{0.5em}
\noindent \textbf{MinRank Problem:}\\
For $\mathbb{F}_q$-linear (matrix) codes, the MinRank problem is stated as follows. Given matrices 
\[
G_1, \dots, G_k \in \mathbb{F}_q^{m \times n}, 
\]
an integer $t$, and a matrix $R \in \mathbb{F}_q^{m \times n}$, the goal is to find a matrix 
\[
E \in \mathbb{F}_q^{m \times n}
\]
of rank at most $t$, and scalars $\lambda_1, \dots, \lambda_k \in \mathbb{F}_q$, such that
\[
R = \lambda_1 G_1 + \cdots + \lambda_k G_k + E.
\]
This problem, which is equivalent to the decoding problem in the rank metric for matrix codes, is known to be NP-complete.

In summary, the Decoding Problem (DP), the Syndrome Decoding Problem (SDP), and the Given Weight Codeword Problem (GWCP) are shown to be equivalent formulations central to algebraic coding theory. These problems underpin many cryptographic constructions based on error-correcting codes. Additionally, several code equivalence problems (including PEP, LEP, and SEP/PKP) extend these ideas, while the rank-metric analogues such as the Rank SDP and the NP-complete MinRank problem further illustrate the complexity landscape in coding theory.

\subsection{Frameworks in Code-Based Cryptography}
This section describes the three existing cryptosystems  built on NP-hard problems in coding theory. We will discuss the McEleice framework, the Niederiter Framework and two variants of the Alekhnovich Framework here. In addition to  these, there are the Quasi-cyclic framework \cite{aguilar2018efficient}, Augot-Finiasz (AF) cryptosystem \cite{10.1007/3-540-39200-9_14} and the GPT cryptosystem \cite{gabidulin1991ideals} which are widely in use and are  derivative variations of the above three frameworks. Details of all the code-based frameworks can be found in \cite{weger2022survey}.

\subsubsection{McEliece Framework}
McEliece \cite{mceliece1978public} proposed a public key cryptosystem using a linear code, for example, a binary Goppa code. The secret key is chosen as one of the many possible generators of a chosen linear code. The public key is a new matrix created by adding randomness and permuting the generator. The new matrix looks like a random matrix and will not leak information about the actual generator. A sender will multiply this random matrix with the plaintext and the product will be added to a random binary error vector whose weight will be less than the error-correcting bound of the chosen linear code. The resulting ciphertext is sent back to the owner of the secret key. To decode the ciphertext, the receiver first performs the decoding algorithm to remove the error vector added in the encoding, then performs the inverse of the operations done to generate the random matrix and retrieve the original message. Without knowledge of the generator and the random transformation, an attacker observing the ciphertext will find it computationally hard to find the message. 

The McEliece framework is described in Table~\ref{tab:mceliece}. The parameters of the framework are $(q, n, k, t)$ where $q$ is a prime or prime power, $n$  the length of the codeword, $k$   the length of the plaintext and $t$   the error correcting capacity of the code. $GL_k$ is the general linear group of order $k$. 

\begin{table}[hbt!]
\centering
\caption{The McEliece Cryptosystem Framework, giving the processes of key generation, encryption and decryption.}
\label{tab:mceliece}
\begin{tabular}{|p{6cm}|p{2.5cm}|p{3.5cm}|}
\hline
\textbf{Transmitter (Key Owner)} & \textbf{Communication} & \textbf{Sender (Plaintext owner)} \\ 
\hline
\multicolumn{3}{|c|}{\textbf{Key Generation}} \\ 
\hline
1. Select linear code \( C \subseteq \mathbb{F}_q^n \) with parameters \([n, k, t]\) & &\\
2. Choose \( k \times n \) generator matrix \( \mathbf{G} \) for \( C \) & &\\
3. Generate a random \(k \times k \) invertible matrix \( \mathbf{S} \in \mathrm{GL}_k(\mathbb{F}_q) \) & &\\
4. Generate a random \( n \times n \) permutation matrix \( \mathbf{P} \)& & \\
5. Compute the Public Key: \( \mathbf{G'} = \mathbf{SGP} \) &  & \\
\textbf{Private Key}: \( (\mathbf{G}, \mathbf{S}, \mathbf{P}) \) & 
\(\xrightarrow{\text{Public Key } (t, \mathbf{G'})}\) & 
 Store public key \( (t, \mathbf{G'}) \) \\ 
\hline
\multicolumn{3}{|c|}{\textbf{Encryption}} \\ 
\hline
~ & ~ & 
1. Encode message \( \mathbf{m} \in \mathbb{F}_q^k \) \\
~ & ~ & 
2. Generate error vector \( \mathbf{e} \in \mathbb{F}_q^n \) with \( \mathrm{wt}(\mathbf{e}) \leq t \) \\
~ & ~ & 
3. Compute ciphertext: \( \mathbf{c} = \mathbf{mG'} + \mathbf{e} \) \\
~ & 
\(\xleftarrow{\text{Ciphertext } \mathbf{c}}\) & ~ \\ 
\hline
\multicolumn{3}{|c|}{\textbf{Decryption}} \\ 
\hline
1. Compute \( \mathbf{cP}^{-1} = \mathbf{mSG} + \mathbf{eP}^{-1} \) & ~ &~ \\
2. Decode using the Private Key \( C \) to recover \( \mathbf{mS} \)  & ~ &~ \\
3. Compute \( \mathbf{m} = \mathbf{mS} \cdot \mathbf{S}^{-1} \) & ~ & ~\\
\hline
\end{tabular}
\end{table}

\subsubsection{ Niederreiter Framework}
The Niederreiter framework \cite{niederreiter1986knapsack} uses the parity-check matrix instead of the generator matrix, resulting in an equivalently secure system. Niederreiter originally proposed using GRS codes as secret codes. In this scheme the plaintext is encoded as a vector of length $n$ and weight less than or equal to the error correcting capacity of the code. The general Niederreiter framework with a linear code is presented in Table ~\ref{tab:niederreiter}.
\begin{table}[hbt!]
\centering
\caption{The Niederreiter Cryptosystem Framework giving the processes of key generation, encryption and decryption.}
\label{tab:niederreiter}
\begin{tabular}{|p{6cm}|p{2.5cm}|p{3.5cm}|}
\hline
\textbf{Transmitter (Key Owner)} & \textbf{Communication} & \textbf{Sender(Plaintext owner)} \\ 
\hline
\multicolumn{3}{|c|}{\textbf{Key Generation}} \\ 
\hline
1. Select a linear code \( C \subseteq \mathbb{F}_q^n \) with parameters \([n, k, t]\) & ~ & ~ \\
2. Choose an \( (n - k) \times n \) parity-check matrix \( \mathbf{H} \) for \( C \) & ~ & ~ \\
3. Generate a random invertible matrix \( \mathbf{S} \in \mathrm{GL}_{n-k}(\mathbb{F}_q) \) & ~ & ~\\
4. Generate a random \( n \times n \) permutation matrix \( \mathbf{P} \) & ~ & ~\\
5. Compute the Public Key: \( \mathbf{H'} = \mathbf{SHP} \) & ~ & ~ \\
\textbf{Private Key}: \( (\mathbf{H}, \mathbf{S}, \mathbf{P}) \) & 
\(\xrightarrow{\text{Public Key } (t, \mathbf{H'})}\) & 
Store public key \( (t, \mathbf{H'}) \) \\ 
\hline
\multicolumn{3}{|c|}{\textbf{Encryption}} \\ 
\hline
~ & ~ & 
1. Encode message \( \mathbf{m} \in \mathbb{F}_q^n \) with \( \mathrm{wt}(\mathbf{m}) \leq t \) \\
~ & ~ & 
2. Compute ciphertext: \( \mathbf{c}^T = \mathbf{H'}\mathbf{m}^T \) \\
~ & 
\(\xleftarrow{\text{Ciphertext } \mathbf{c}}\) & ~ \\ 
\hline
\multicolumn{3}{|c|}{\textbf{Decryption}} \\ 
\hline
1. Compute \( \mathbf{S}^{-1}\mathbf{c}^T = \mathbf{HP}\mathbf{m}^T \) & ~ & ~\\
2. Use the decoding algorithm of \( C \) to recover \( \mathbf{P}\mathbf{m}^T \) & ~ & ~\\
3. Compute \( \mathbf{m}^T = \mathbf{P}^{-1}(\mathbf{P}\mathbf{m}^T) \) & ~ & ~ \\ 
\hline
\end{tabular}
\end{table}

\subsubsection{Alekhnovich's Cryptosystem}
Alekhnovich's cryptosystem \cite{alekhnovich2003more} is the first code-based cryptosystem with provable security proof. It relies solely on the decoding problem and lays the foundations for modern code-based cryptography, where researchers attempt to construct code-based cryptosystems with a provable reduction to the problem of distinguishing a random codeword from a uniform string. There are two variants of this cryptosystem, both relying on the hard problem of distinguishing a random vector from an erroneous codeword of the code $C$. We present the first (encrypts one bit at a time) and second (encrypts a vector of bits at a time) variants of the Alkehnovich cryptosystem in Tables \ref{tab:alekhnovich} and \ref{tab:alekhnovich_second}. In both variants $n$ is the length of the ciphertext, $t$  the error-correcting bound of $C$ and $k$  the number of rows in the generator matrix of $C$. 

\begin{table}[hbt!]
\centering
\caption{The Alekhnovich First Variant Framework, giving the processes of key generation, encryption and decryption.}
\label{tab:alekhnovich}
\begin{tabular}{|p{6cm}|p{2.5cm}|p{3.5cm}|}
\hline
\textbf{Transmitter (Key Owner)} & \textbf{Communication} & \textbf{Sender (Plaintext owner)} \\ 
\hline
\multicolumn{3}{|c|}{\textbf{Key Generation}} \\ 
\hline
1. Select parameters: \( t \in o(\sqrt{n}) \) & ~ & ~\\
2. Choose random matrix \( \mathbf{A} \in \mathbb{F}_2^{k \times n} \) & ~ & ~\\
3. Generate random vector \( \mathbf{e} \in \mathbb{F}_2^n \) with \( \mathrm{wt}(\mathbf{e}) = t \) & ~ & ~\\
4. Generate random vector \( \mathbf{x} \in \mathbb{F}_2^k \) & ~ & ~\\
5. Compute \( \mathbf{y} = \mathbf{xA} + \mathbf{e} \) & ~ & ~\\
6. Construct \( \mathbf{H}^T = (\mathbf{A}^T, \mathbf{y}^T) \)& ~ & ~ \\
7. Define code \( C = \ker(\mathbf{H}) \) & ~ & ~\\
8. Choose generator matrix \( \mathbf{G} \in \mathbb{F}_2^{(n-k) \times n} \) for \( C \) & ~ & ~\\
\textbf{Public Key}: \( (\mathbf{G}, t) \)& ~ & ~ \\
\textbf{Private Key}: \( \mathbf{e} \) & 
\(\xrightarrow{\text{Public Key } (\mathbf{G}, t)}\) & 
Store public key \( (\mathbf{G}, t) \)  \\ 
\hline
\multicolumn{3}{|c|}{\textbf{Encryption}} \\ 
\hline
~ & ~ & 
1. Encode message \( \mathbf{m} \in \mathbb{F}_2 \) \\
~ & ~ & 
2. If \( \mathbf{m} = 0 \): \\
~ & ~ & \quad - Choose \( \mathbf{a} \in \mathbb{F}_2^{n-k} \) \\
~ & ~ & \quad - Choose \( \mathbf{e'} \in \mathbb{F}_2^n \) with \( \mathrm{wt}(\mathbf{e'}) = t \) \\
~ & ~ & \quad - Compute ciphertext: \( \mathbf{c} = \mathbf{aG} + \mathbf{e'} \) \\
~ & ~ & 
3. If \( \mathbf{m} = 1 \): \\
~ & ~ & \quad - Choose random \( \mathbf{c} \in \mathbb{F}_2^n \) \\
~ & 
\(\xleftarrow{\text{Ciphertext } \mathbf{c}}\) & ~ \\ 
\hline
\multicolumn{3}{|c|}{\textbf{Decryption}} \\ 
\hline
1. Compute \( \mathbf{b} = \langle \mathbf{e}, \mathbf{c} \rangle \) & ~ & ~\\
2. If \( \mathbf{b} = 0 \): Decrypt \( \mathbf{m} = 0 \) (high probability) & ~ & ~\\
3. If \( \mathbf{b} = 1 \): Decrypt \( \mathbf{m} = 1 \) (probability \( 1/2 \)) & ~ & ~ \\ 
\hline
\end{tabular}
\end{table}

\begin{table}[hbt!]
\centering
\caption{The Alekhnovich Second Variant Framework, giving the processes of key generation, encryption and decryption.}
\label{tab:alekhnovich_second}
\begin{tabular}{|p{6cm}|p{2.5cm}|p{3.5cm}|}
\hline
\textbf{Transmitter (Key Owner)} & \textbf{Communication} & \textbf{Sender (Plaintext owner)} \\ 
\hline
\multicolumn{3}{|c|}{\textbf{Key Generation}} \\ 
\hline
1. Choose random matrices:& ~ & ~ \\
\quad - \( \mathbf{A} \in \mathbb{F}_2^{n/2 \times n} \) & ~ & ~\\
\quad - \( \mathbf{X} \in \mathbb{F}_2^{n \times n/2} \) & ~ & ~\\
\quad - \( \mathbf{E} \in \mathbb{F}_2^{n \times n} \) with row weight \( t \) & ~ & ~\\
2. Compute \( \mathbf{M} = \mathbf{XA} + \mathbf{E} \in \mathrm{GL}_n(\mathbb{F}_2) \) & ~ & ~\\
3. Define binary code \( C_0 \) capable of correcting errors from a BSC with transition probability \( t^2/n \) & ~ & ~\\
4. Define map \( \phi: \mathbf{x} \mapsto \mathbf{Mx} \) & ~ & ~\\
5. Construct code \( C = \phi^{-1}(C_0) \cap \ker(\mathbf{A}) \)& ~ & ~ \\
6. Choose generator matrix \( \mathbf{G} \in \mathbb{F}_2^{k \times n} \) for \( C \)& ~ & ~ \\
\textbf{Public Key}: \( (\mathbf{G}, t) \)& ~ & ~ \\
\textbf{Private Key}: \( \mathbf{E} \) & 
\(\xrightarrow{\text{Public Key } (\mathbf{G}, t)}\) & 
Store public key \( (\mathbf{G}, t) \) \\ 
\hline
\multicolumn{3}{|c|}{\textbf{Encryption}} \\ 
\hline
~ & ~ & 
1. Encode message \( \mathbf{m} \in \mathbb{F}_2^{k/2} \) \\
~ & ~ & 
2. Choose random \( \mathbf{r} \in \mathbb{F}_2^{k/2} \) \\
~ & ~ & 
3. Choose random \( \mathbf{e} \in \mathbb{F}_2^n \) with \( \mathrm{wt}(\mathbf{e}) = t \) \\
~ & ~ & 
4. Compute \( \mathbf{x} = (\mathbf{m}, \mathbf{r}) \in \mathbb{F}_2^k \) \\
~ & ~ & 
5. Compute ciphertext: \( \mathbf{c} = \mathbf{xG} + \mathbf{e} \) \\
~ & 
\(\xleftarrow{\text{Ciphertext } \mathbf{c}}\) & ~ \\ 
\hline
\multicolumn{3}{|c|}{\textbf{Decryption}} \\ 
\hline
1. Compute \( \mathbf{y}^T = \mathbf{E}\mathbf{c}^T = \mathbf{z}^T + \mathbf{E}\mathbf{e}^T \)& ~ & ~ \\
2. Decode \( \mathbf{y} \) using \( C_0 \) to recover \( \mathbf{z} \) & ~ & ~\\
3. Solve \( \mathbf{xG} = \phi^{-1}(\mathbf{z}) \) to recover \( \mathbf{x} \)& ~ & ~ \\
4. Extract message \( \mathbf{m} \) from \( \mathbf{x} \) & ~ & ~ \\ 
\hline
\end{tabular}
\end{table}
\par

Despite the security advantages of code-based ciphers the challenges that have impeded its practical application are the large size of the keys and ciphertext expansion, making it costly and inefficient to communicate.


\subsection{Advantages in the Post-Quantum Era}
Code-based cryptosystems are known for their strong resistance to quantum attacks as no known quantum algorithms efficiently solve SDP or distinguish Goppa codes. Additionally, they involve simpler arithmetic operations compared to lattice-based schemes and avoid complex noise management techniques. Their relative simplicity and efficiency in implementation compared to lattice-based schemes make them attractive candidates for post-quantum cryptographic applications.

\begin{itemize}
    \item \textbf{Quantum Resistance}: 
    Code-based cryptography derives its security from the \textit{Syndrome Decoding Problem (SDP)}, which has been proven NP-hard \cite{barg1998complexity}, and the \textit{Goppa Code Distinguishing Problem}. Unlike the Shor-vulnerable RSA/ECC systems, no known quantum algorithm solves SDP sub-exponentially \cite{berstein2017classic}. The Classic McEliece cryptosystem, a NIST Post-Quantum Standardization finalist \cite{nist_pqc_2022}, exemplifies this resilience. While lattice-based schemes rely on LWE/Ring-LWE assumptions vulnerable to future quantum advances \cite{albrecht2018lwe}, code-based systems maintain security through decades-old coding theory foundations \cite{sendrier2007structural}.

    \item \textbf{Efficiency}: 
    Code-based HE replaces lattice-based polynomial ring arithmetic with finite field matrix operations. For example, encryption in McEliece variants requires only $\mathcal{O}(n^2)$ matrix-vector multiplications, compared to $\mathcal{O}(n^3)$ FFT-accelerated polynomial multiplications in Ring-LWE schemes \cite{martinez2022code_based_rlwe}. Martínez et al.\cite{martinez2022code_based_rlwe} demonstrated 1.8$\times$ faster encryption in code-based RLWE hybrids compared to pure lattice implementations. The absence of probabilistic decryption failures further reduces redundant computations \cite{dottling2019code_based}.

    \item \textbf{Implementation Simplicity}: 
    Code-based schemes inherently manage noise through error-correcting codes rather than artificial noise sampling. Armknecht et al. \cite{armknecht2011constructing} showed this eliminates lattice-style noise flooding and modulus switching. For example, McEliece-based HE uses predetermined error vectors from code distance properties, avoiding Gentry's \cite{gentry2009fully} complex bootstrapping framework. Recent implementations by Chen et al. \cite{chen2023homomorphic} required 40\% fewer code lines compared to lattice-based libraries. This simplicity extends to hardware acceleration - FPGA implementations show 
    2.3$\times$ better area-time product than lattice analogs \cite{guo2021hardware}.
\end{itemize}

\section{Code-Based Homomorphic Encryption Schemes} \label{Sec:CBHES}
This section presents  code-based HE constructions, along with their functionality and comparative analysis.
\subsection{Bogdanov and Lee Homomorphic Encryption}
This scheme\cite{bogdanov2011homomorphic} was constructed by combining the encryption structure
of the local cryptosystem of Applebaum, Barak, and Wigderson \cite{applebaum2010public} with a key scrambling of the McEliece cryptosystem \cite{mceliece1978public}.\\
The ABW PKE scheme is based on hardness-on-average assumptions for natural combinatorial NP-hard optimization problems with the following assumptions:
\begin{enumerate}
    \item  It is infeasible to solve a random set of sparse linear equations mod 2, of which a small fraction is noisy.
    \item It is infeasible to distinguish between a random unbalanced bipartite graph and a graph in which a set $S$ with only $|S|/3$ neighbors is planted at random on the large side.
\item There is a pseudo-random generator, where every output depends on a random subset of the constant size of the input.
\end{enumerate}

 The basic idea here is to construct a generator matrix for the McEliece cryptosystem with the above assumptions.
 The encryption scheme can be described below:
 \begin{enumerate}
     \item \textbf{KeyGen}\\
 Choose a uniformly random subset $S' \subseteq \{1, \cdots ,n\}$ of size $s$ and an $n \times r$ matrix
$M$ from the following distribution. First, choose a set of uniformly random but distinct values
$a_1, \cdots, a_n$ from $F_q$. Set the $i^{th}$ row $M_i$ to

\begin{equation*}
 M_i = 
\left\{
    \begin{array}{lr}
        (a_i a_i^2 \cdots a_i^{s/3} 0 \cdots 0), & \text{if } i \in S' \\
        (a_i a_i^2 \cdots a_i^{s/3} a_i^{s/3+1} \cdots a_i^r), & \text{if } i \notin S'
    \end{array}
\right\}
\end{equation*}


The secret key is the pair $(S', M)$ and the public key is the matrix $P = MR$, where $R$ is a random $r \times r$ matrix over $F_q$ with determinant 1.
\item \textbf{Encryption} \\
Given a public key $P$, to encrypt a message $m \in F_q$, choose a uniformly random $x \in F_q^r$ and an error vector $e \in F_q^n$ by choosing each of its entries independently at random from a random distribution $\chi$. Output the ciphertext $c=Px + m1 + e$, where $1 \in F_q^n$ is the all ones vector.
\item \textbf{Decryption}\\
 Given a secret key $(S', M)$, to decrypt a ciphertext $c \in F_q^n$, first find a solution to the
following system of $(s/3) + 1$ linear equations over variables $y_i \in F_q, i \in S'$
$$\sum_{i \in S'} y_iM_i=0,$$
$$\sum_{i \in S'}y_i=1,$$
with $y_i = 0$ when $i \notin S'$. Output the value $\sum_{i\in[n]}y_ic_i$.
 \end{enumerate}
\subsection{Armknecht Scheme}
Armknecht \cite{armknecht2011constructing} provided a generic construction of a symmetric key homomorphic encryption scheme that can evaluate multivariate polynomials up to a fixed degree $\mu$.
\begin{enumerate}
    \item \textbf{Keygen(s,$\mu$, L)}\\
    The input $s$ represents the security parameter, $L$ is the expected total number of encryptions, and $\mu$ is the maximum degree of supported polynomials. The setup algorithm will then select a codeword support $x$, a message support $y$, and two special evaluation codes $C$ and $C'$ in such a way that $C^\mu \subseteq C'$, and the length of codewords is at least L. The choice of appropriate codes and parameters will vary depending on the coding scheme. Keygen will generate a set, $I$, of size T and a subset of $[n]$, where $[n]$ is the set \(\{1,2 \cdots n\}\). T depends on the above parameter and the deployed code. I represents the good locations for the generated encryptions and serves as the secret key of the scheme. The final output will be the secret key $k = (x, y, I)$.
    \item \textbf{Encrypt(m,k)}\\
    The inputs are a plaintext message  $m \in \mathbb{F}$, and a secret key
$k = (x, y, I)$. Encrypt first chooses a random encoding $w \in C$ of m, using the Encode algorithm and the knowledge of the supports $x$ and $y$. Then, it samples a uniformly random error vector $e \in \mathbb{F}^n$, such that $supp(e) \subseteq [n] \ I$ and computes $c = w + e$.
Finally, the ciphertext is defined as the pair $(c, 1)$ where the first entry is an erroneous codeword in $C(I)$ that encodes the plaintext m while the second entry, the integer, is a counter to keep track of the number of multiplications.
\item \textbf{Decrypt($(c,\gamma),k)$)}\\
Decrypt gets as input the secret key $k = (x,y, I)$ and a pair $(c, \gamma)$ with $c \in C(I)$ and $\gamma \leq \mu$. It outputs $m = Decode(c, I)$ where Decode is used with respect to $x$ and $y$.
\item \textbf{Add($(c_1,\gamma_1),(c_2,\gamma_2)$)},
 outputs $(c_1+c_2,\max(\gamma_1,\gamma_2))$.
 \item \textbf{Mult($(c_1,\gamma_1),(c_2,\gamma_2)$},
 outputs $(c_1 \cdot c_2,\gamma_1+\gamma_2)$.
\end{enumerate}

Next, we will discuss Algebraic Geometry (AG)
codes and types in detail for the applications of this scheme. We will also compare the variations of the codes to be used in this scheme. 
\subsubsection{Comparison and Security Implications from an AG Code Perspective}

AG codes are codes constructed from algebraic curves over finite fields, providing a flexible framework for designing codes with various parameters and properties. From the perspective of AG codes, we can compare Reed-Muller (RM) codes, Reed-Solomon (RS) codes, and Goppa codes as follows:

\textbf{Reed-Muller Codes:} RM codes can be regarded as a special case of AG codes, specifically constructed from the projective line. They have a relatively simple structure, making them easier to analyze.

\textbf{Reed-Solomon Codes:} RS codes are also a special case of AG codes, but they are constructed from the projective line in a different manner. They exhibit a more complex structure compared to RM codes and offer enhanced error-correcting capabilities.

\textbf{Goppa Codes:} Goppa codes are more general than both RM and RS codes, as they can be constructed from a wider array of algebraic curves. This flexibility allows for the design of codes with improved security properties and higher error-correcting capabilities.

Goppa codes are generally considered to be more secure than RM and RS codes due to their less understood structure, along with the ability to choose parameters that enhance resistance to known attacks. However, it is essential to note that the security of any code-based cryptosystem relies on the specific choice of parameters and the use of appropriate decoding algorithms. A comparison of RM, RS and Goppa Codes can be found in Table \ref{tab:code}.

\begin{table}[ht]
\caption{Comparison of the three different Alegbraic Geometry Codes detailed above.}
    \begin{tabular}{|l|p{3.5cm}|p{3.5cm}|p{3.5cm}|}
\hline
--- & \textbf{Reed-Muller (RM) Codes} & \textbf{Reed-Solomon (RS) Codes} & \textbf{Goppa Codes} \\
\hline 
Definition & Linear code defined by parity check matrix of monomials & Linear code defined by parity check matrix of polynomials & Linear code defined by a Goppa polynomial and a set of elements \\
\hline
Encoding & Multiplication by generator matrix & Multiplication by generator matrix & Multiplication by generator matrix \\
\hline
Decoding & Majority-logic decoding, Berlekamp's algorithm & Berlekamp-Massey decoding, Euclidean algorithm & Sudan-Guruswami algorithm, McEliece cryptosystem \\
\hline
Efficiency & Generally less efficient than RS and Goppa codes & Efficient encoding and decoding & Efficient decoding for certain parameters \\
\hline
Security & Less secure than Goppa codes due to well-understood structure & Moderate security & Generally considered more secure than RM and RS codes \\
\hline
Key Size & Larger than RS codes for the same level of security & Smaller than RM and Goppa codes & Can be larger than RS codes depending on parameters \\
\hline
\end{tabular}

\label{tab:code}
\end{table}

\subsection{Rank Metric Based Homomorphic Encryption}
The rank-metric-based homomorphic encryption scheme, introduced by Aguilar et al. \cite{aguilar2025somewhat}, marks a significant advancement in the field of homomorphic encryption. The authors initially developed an additively homomorphic encryption (AHE) scheme and later extended it to support multiplicative homomorphism. By incorporating bootstrapping techniques and various optimizations, they transformed the scheme into a somewhat homomorphic encryption (SWHE) system capable of performing both addition and multiplication on encrypted data. 

The AHE scheme is parameterized by the following values:
\begin{itemize}
    \item \( q \): The cardinality of the base field.
    \item \( m \): The dimension of the field extension.
    \item \( n \): The length of the vectors.
    \item \( w \): The rank weight of the error, where \( w < m \).
\end{itemize}

\textbf{Key Generation (KeyGenAHE)}

The key generation algorithm proceeds as follows:
\begin{enumerate}
    \item Sample \( f = (f_1, \dots, f_w) \xleftarrow{\$} S_w(\mathbb{F}_{q^m}) \), where \( S_w(\mathbb{F}_{q^m}) \) denotes the set of \( w \)-dimensional subspaces over \( \mathbb{F}_{q^m} \).
    \item Extend \( f \) into a basis \( b = (f_1, \dots, f_w, g_1, \dots, g_{m-w}) \in S_m(\mathbb{F}_{q^m}) \).
    \item Define \( g = (g_1, \dots, g_{m-w}) \).
    \item Compute the matrix \( B = \text{Mat}(b) \), by extending each element of $b$ from \(\mathbb{F}_{q^m}\) to \(\mathbb{F}_{q}\).
    \item Define \( D \) as the last \( m-w \) columns of \( (B^{-1})^T \).
    \item Sample \( s \xleftarrow{\$} \mathbb{F}_q^n \) with \( F = \text{supp}(f) \).
    \item Return the secret key \( sk = (f, g, D, s) \).
\end{enumerate}

 \textbf{Encryption (EncryptAHE)}

The encryption algorithm is defined as follows:
\begin{enumerate}
    \item Sample \( r = (r_1, R_2) \xleftarrow{\$} \mathbb{F}_{q^m}^n \times M_{w,n}(\mathbb{F}_q) \), where \( M_{w,n}(\mathbb{F}_q) \) denotes the set of \( w \times n \) matrices over \( \mathbb{F}_q \).
    \item Compute \( u = r_1 \) and \( e = fR_2 \).
    \item Compute \( v = s \cdot u + e + \hat{m} \), where \( \hat{m} = g(1) \star m \in \mathbb{F}_{q^m}^n \), and  \(g(1)\) is the first row of \(g\).
    \item Return the ciphertext \( ct = (u, v) \).
\end{enumerate}

\textbf{Decryption (DecryptAHE)}

The decryption algorithm proceeds as follows:
\begin{enumerate}
    \item Compute \( d^T \text{Mat}(v - s \cdot u) \), where \( d = D(1) \).
\end{enumerate}

The EncryptAHE and DecryptAHE algorithms are deterministic functions, meaning their outputs are solely determined by their inputs. The randomness in the encryption process is incorporated within the EncryptAHE function itself, ensuring that the function remains deterministic given the input message and the randomly sampled values. This property is crucial for the correctness and security of the AHE scheme.

In the EncryptAHE algorithm, the error term \( e \) is sampled as \( e = fR_2 \), where \( R_2 \) is randomly sampled from the matrix space \( M_{w,n}(\mathbb{F}_q) \). This is mathematically equivalent to directly sampling \( e \) from the vector space \( \mathbb{F}_q^n \). Additionally, the term \( \hat{m} \), derived from the message \( m \) using the basis vector \( g(1) \), is restricted to the subspace \( \langle g(1) \rangle_{\mathbb{F}_q}^n \). This restriction plays a crucial role in ensuring the correctness and security of the AHE scheme.

To extend the AHE scheme to support multiplicative homomorphism, the following algorithms are introduced:

Given two ciphertexts \( ct = (u, v) \) and \( ct' = (u', v') \), the multiplication algorithm computes:
\[
\text{Mul}(ct, ct') = (v \cdot v', -(u \cdot v' + u' \cdot v), u \cdot u') \in \mathbb{F}_{q^m}^n \times \mathbb{F}_{q^m}^n \times \mathbb{F}_{q^m}^n.
\]

Given the secret key \( sk = (f, g, D, s) \) and the resulting ciphertext \( (a, b, c) \) from the multiplication operation, the decryption algorithm computes:
\[
\text{DecryptMul}(sk, (a, b, c)) = d^T \text{Mat}(a + s \cdot b + s \cdot s \cdot c),
\]
where \( d = D(2) \), the second column of D .

Key switching is a technique used to change the encryption key while preserving the ciphertext. This is essential for refreshing the key to improve security or to enable more efficient operations.

The key generation algorithm for key switching proceeds as follows:
\begin{enumerate}
    \item Generate a new basis \( b_2 \) of \( \mathbb{F}_{q^m} \) and a new secret key \( sk_2 = (f_2, g_2, D_2, s_2) \) as in KeyGenSHE.
    \item For \( 1 \leq i \leq m \), define \( s_{1,i} = d^T \text{Mat}(\gamma_i \star s_1) \), where \( d = D(1) \), the first column of D.
    \item Define \( ksk = (ksk_1, \dots, ksk_m) \), where \( ksk_i \xleftarrow{\$} \text{EncryptSHE}(sk_2, s_{1,i}) \).
    \item For \( 1 \leq i \leq m \), define \( p_i = d(i) \star (1, 0, \dots, 0) \in \mathbb{F}_{q^m}^n \), where \(d(i)\) is the $i^{th}$ column of $D$.
    \item Define \( projk = (projk_1, \dots, projk_m) \), where \( projk_i \xleftarrow{\$} \text{EncryptSHE}(sk_2, p_i) \).
    \item Return \( (sk_2, ksk, projk) \).
\end{enumerate}

The homomorphic decryption algorithm with key switching proceeds as follows:
\begin{enumerate}
    \item Define \( u_i \in \mathbb{F}_q^n \) such that \( u = \sum_{i=1}^m \gamma_i \star u_i \).
    \item Define \( v_i \in \mathbb{F}_q^n \) such that \( v = \sum_{i=1}^m \gamma_i \star v_i \).
    \item Compute \( ct_1 = \sum_{i=1}^m v_i \cdot projk_i \).
    \item Compute \( ct_2 = \sum_{i=1}^m u_i \cdot ksk_i \).
    \item Return \( ct_1 - ct_2 \).
\end{enumerate}

The rank-metric-based homomorphic encryption scheme extends the capabilities of additively homomorphic encryption to support multiplicative homomorphism through the introduction of multiplication and key-switching algorithms. These enhancements enable the scheme to perform basic computations on encrypted data, making it suitable for applications requiring privacy-preserving computations. The deterministic nature of the encryption and decryption processes, combined with careful error sampling and message encoding, ensures the correctness and security of the scheme. Key switching further enhances the flexibility and security of the system, allowing for efficient key management and homomorphic operations.

\subsection{Comparison of Code-based HE with Lattice-Based HE}
As illustrated in Table \ref{tab:efficiency}, code‐based HE schemes typically suffer from significantly larger key sizes—often in the order of megabytes or even gigabytes—due to their reliance on matrix-based structures such as those in the McEliece cryptosystem. While their operations (often based on simple XOR and matrix-vector multiplications) can be parallelized, the computational overhead remains high. In contrast, lattice‐based HE schemes, particularly those built on Ring-LWE, benefit from optimizations like the Number Theoretic Transform (NTT) that allow for much more compact key representations (usually in the range of 1–10 KB) and faster arithmetic operations. Ciphertext expansion in code-based schemes tends to be high. In contrast, lattice-based schemes offer more moderate expansion—for example, schemes like CKKS typically expand ciphertexts by about $8\times$ relative to the plaintext.

\begin{table}[H]
\centering
\caption{Efficiency Comparison of code-based vs lattice-based Homomorphic encryption.}
\label{tab:efficiency}
\begin{tabular}{|p{3cm}|p{4cm}|p{4cm}|}
\hline
\textbf{Metric} & \textbf{Code-Based HE} & \textbf{Lattice-Based HE} \\
\hline
\textbf{Key Sizes} & Large (e.g., McEliece keys: $\approx$1 MB--1 GB due to matrix-based structures). Recent code-based FHE schemes still struggle with key size reduction. & Smaller (e.g., Ring-LWE keys: $\approx$ 1-10 KB). Optimizations like NTT (Number Theoretic Transform) enable compact representations. \\
\hline
\textbf{Computation Speed} & Matrix/vector operations are parallelizable but computationally heavy. Simpler arithmetic (e.g., XOR-based operations in some schemes). & Faster due to polynomial ring optimizations (e.g., NTT accelerates multiplication in CKKS/BGV \cite{cheon2017homomorphic}). \\
\hline
\textbf{Ciphertext Expansion} & High (e.g., ciphertexts are large matrices or vectors). & Moderate (e.g., CKKS ciphertexts expand ~8x plaintext size). \\
\hline
\end{tabular}
\end{table}

Noise management is a critical aspect of HE schemes. Table \ref{tab:noise} summarizes that code-based HE inherits inherent error correction from the underlying codes (such as Goppa codes), which helps to keep noise growth bounded in additive settings. However, noise growth in these schemes is generally linear, as seen in some additive constructions \cite{armknecht2011constructing}. In lattice-based HE, the noise grows polynomially with each multiplicative operation, necessitating frequent noise management strategies modulus switching and bootstrapping. Mature bootstrapping techniques, as pioneered by Gentry and further optimized in schemes such as BGV \cite{brakerski2012fully}, mitigate the effects of noise accumulation, albeit at an additional computational cost.

\begin{table}[H]
\centering
\caption{Noise Management Comparison of code-based vs lattice-based homomorphic encryption.}
\label{tab:noise}
\begin{tabular}{|p{4cm}|p{4cm}|p{4cm}|}
\hline
\textbf{Aspect} & \textbf{Code-Based HE} & \textbf{Lattice-Based HE} \\
\hline
\textbf{Noise Growth} & Inherits error-correction properties: errors are intentionally added but bounded by code distance. Noise grows linearly in some additive schemes (e.g., \cite{armknecht2011constructing}). & Noise grows polynomially with multiplicative operations. Requires frequent management (e.g., modulus switching). \\
\hline
\textbf{Bootstrapping} & Limited progress, with high overhead. & Mature techniques (e.g., Gentry’s bootstrapping in BGV \cite{brakerski2012fully}). Optimized via sparse embeddings or hybrid key-switching. \\
\hline
\textbf{Error Correction} & Built-in error correction can mitigate noise. & Relies on probabilistic decryption; no inherent error correction. \\
\hline
\end{tabular}
\end{table}

The security foundations of code-based HE rest on NP-hard problems like the Syndrome Decoding Problem (SDP) and the Linear Code Distinguishing Problem. These problems have withstood quantum attacks, as there is no known quantum speedup for solving them. However, the large key sizes and relatively immature FHE implementations have limited their widespread adoption (Table \ref{tab:security}). On the other hand, Lattice-based HE relies on newer assumptions such as Learning With Errors (LWE) and its ring variant (RLWE), which provide strong security guarantees via worst-case hardness reductions. While these schemes have been integrated into many NIST-backed PQC standards and benefit from extensive industrial testing, their security also depends on emerging lattice reduction techniques and careful management of side-channel vulnerabilities.

\begin{table}[H]
\centering
\caption{Security Assumptions and Trade-offs between code-based vs lattice-based homomorphic encryption.}
\label{tab:security}
\begin{tabular}{|p{4cm}|p{4cm}|p{4cm}|}
\hline
\textbf{Criterion} & \textbf{Code-Based HE} & \textbf{Lattice-Based HE} \\
\hline
\textbf{Core Hard Problems} & Syndrome Decoding Problem (SDP), Linear Code Distinguishing Problem (NP-hard). & Learning With Errors (LWE), Ring-LWE (reductions to worst-case lattice problems). \\
\hline
\textbf{Quantum Resistance} & SDP has no known quantum speedup; robust against Shor’s/Grover’s algorithms. & LWE/Ring-LWE are quantum-resistant but rely on newer assumptions. \\
\hline
\textbf{Standardization} & Limited adoption (e.g., Classic McEliece is a NIST PQC finalist but not HE-focused). & Dominates NIST PQC standards (e.g., Kyber, Dilithium). HE schemes (CKKS/BGV) are industry-tested. \\
\hline
\textbf{Attack Surface} & Structural attacks (e.g., weak code choice) and ISD attacks (exponential time). & Side-channel attacks, decryption failures, and novel lattice reductions (e.g., BKZ). \\
\hline
\end{tabular}
\end{table}

\begin{table}[H]
\centering
\caption{Summary of Trade-offs between code-based vs lattice-based homomorphic encryption.}
\label{tab:tradeoffs}
\begin{tabular}{|p{4cm}|p{4cm}|p{4cm}|}
\hline
\textbf{Category} & \textbf{Code-Based HE} & \textbf{Lattice-Based HE} \\
\hline
\textbf{Strengths} & - Provable NP-hard security.\newline - Built-in error correction.\newline - Simpler operations. & - Efficient bootstrapping.\newline - Optimized libraries (e.g., Microsoft SEAL).\newline - NIST-backed. \\
\hline
\textbf{Weaknesses} & - Large keys/ciphertexts.\newline - Immature FHE implementations. & - Complex noise management.\newline - Relies on newer security assumptions. \\
\hline
\end{tabular}
\end{table}

Table \ref{tab:tradeoffs} provides a concise summary of the strengths and weaknesses of both approaches. Code-based HE offers provable NP-hard security and built-in error correction along with simpler operational models. However, these advantages are offset by practical challenges such as large key and ciphertext sizes and less mature implementations. Lattice-based HE schemes, conversely, enjoy efficient bootstrapping, a host of optimized libraries (e.g., Microsoft SEAL), and robust backing from NIST initiatives. Their primary challenges lie in complex noise management and the reliance on relatively newer security assumptions that continue to be scrutinized by the research community.

The security of homomorphic encryption (HE) schemes, especially those based on lattice and code structures, is an area of ongoing research and evaluation. Lattice-based HE schemes, such as those relying on the Learning With Errors (LWE) problem, have played a crucial role in the development of fully homomorphic encryption systems. However, recent advancements in quantum algorithms have raised concerns about their long-term security. A recent study \cite{chen2024quantum} proposed a polynomial-time quantum algorithm for solving the LWE problem under certain polynomial modulus-noise ratios; if this finding is correct, it could undermine the security assumptions of lattice-based HE schemes. Although a flaw was identified in this algorithm, the search for quantum solutions highlights potential vulnerabilities in lattice-based cryptosystems. On the other hand, code-based HE schemes, while less common, provide alternative security bases rooted in different hard problems, such as the Ideal Rank Syndrome Decoding (IRSD) problem. The security of these schemes is often associated with the difficulty of decoding random linear codes, a challenge that remains difficult even for quantum computers. A code-based somewhat homomorphic encryption scheme \cite{aguilar2025somewhat} has demonstrated security based on the IRSD problem, indicating resilience against certain quantum attacks.

\section{Conclusion and Future Work}\label{Sec:Con}
Code-based homomorphic encryption offers promise in the realm of post-quantum cryptography. Its strong security foundation is based on NP-hard problems, such as the syndrome decoding problem, and includes built-in error-correction capabilities, which provide robust protection against both classical and quantum threats. Although there are current challenges related to efficiency and noise management, ongoing research and efforts toward standardization are enhancing its practical implementation. As the cryptographic community continues to explore and improve these schemes, code-based homomorphic encryption is set to play a crucial role in securing sensitive computations in a future where quantum computing becomes a reality.
\subsection{Challenges in Code-Based HE}
Code-based homomorphic encryption faces several technical challenges that must be addressed to fully harness its potential. Key issues include improving computational efficiency, reducing the overheads associated with large key and ciphertext sizes, and managing noise accumulation. While the inherent error-correcting properties of certain code families provide an advantage in controlling noise, significant work is still needed to refine these schemes for practical, high-performance deployment.

Braverski \cite{braverski2022homomorphism} demonstrated that LPN (Learning Parity with Noise)-based schemes face fundamental limitations when adapted for homomorphic encryption, primarily due to their linear algebraic structure. These limitations directly parallel challenges in code-based HE. LPN-based HE relies on linear operations over finite fields (e.g., XORs). Braverski shows that repeated homomorphic operations expose linear relationships between ciphertexts, enabling adversaries to recover secret keys via linear algebra attacks. Braverski highlights that enforcing homomorphism in LPN requires relaxing noise distributions or code parameters, inadvertently reducing security margins.
Code-based HE (McEliece variants) also depends on linear codes, where ciphertexts are generated via matrix-vector multiplications. Homomorphic operations (e.g., additions) preserve this linearity, creating similar attack surfaces. An adversary could exploit homomorphic additions to derive the structure of the generator matrix, weakening code secrecy \cite{armknecht2011constructing}. So to enforce Homomorphic encryption, a novel framework has to be constructed, which does not rely on algebraic structure.

\subsection{Research Opportunities}
Code-based cryptography offers a promising avenue for post-quantum HE, but several challenges remain before it can be considered a practical standard. Addressing the efficiency of bootstrapping, key size reduction, error management, and security proofs are critical areas of future research. Advancements in these domains will be essential to establish code-based HE as a viable alternative to lattice-based schemes in the post-quantum era.

\begin{enumerate}
    \item  Limited Fully Homomorphic Encryption (FHE) Constructions \\
Most research on post-quantum HE focuses on lattice-based schemes, leaving code-based HE relatively underdeveloped. The main challenge with the latter lies in designing efficient bootstrapping techniques for fully homomorphic encryption (FHE) based on coding theory. Existing proposals, such as those built on the McEliece cryptosystem, do not inherently support homomorphic operations efficiently. Developing bootstrapping methods that enable arbitrary-depth computation while maintaining practical key sizes and computational efficiency remains an open problem \cite{Misoczki2013, Baldi2019}.

\item Key Size and Efficiency Constraints \\
One of the primary barriers to widespread adoption of code-based cryptographic schemes, including HE, is the relatively large key sizes. Traditional code-based encryption, such as the McEliece and Niederreiter cryptosystems, requires public keys that can be several megabytes in size. While variants using Quasi-Cyclic Moderate-Density Parity-Check (QC-MDPC) codes and Rank-Metric codes have been proposed to mitigate this issue, further research is needed to develop compression techniques and structured code families that reduce key sizes without compromising security \cite{Misoczki2013, AguilarMelchor2018}. An alternate approach to tackle this problem can be found in the algebraic structure of the Torus. Chilloti et.al. \cite{chillotti2019tfhe} constructed a third-generation homomorphic encryption scheme on a Torus that proved to be efficient in encrypting the data homomorphically and demonstrated scalablity in operations through modular arithmetic. If the computations on a code-based cryptosystem can be translated from a Field to a Torus, we have an opportunity to achieve efficiency along with the code-based guarantees of security against chosen plaintext attacks and ciphertext-only attacks.

\item Efficient Error-Correction Mechanisms for HE Operations \\
Homomorphic operations in code-based cryptography require mechanisms for handling accumulated noise, similar to error growth in lattice-based schemes. However, unlike LWE-based HE, where noise growth is well-studied, error propagation in code-based HE remains an open area. Designing efficient error-correction techniques that maintain the correctness of homomorphic computations while minimizing decryption failures is a crucial research direction \cite{Baldi2019, Persichetti2020}.

\item Security Assumptions and Standardization \\
While code-based cryptosystems have withstood decades of cryptanalysis, their homomorphic counterparts lack comprehensive security analyses. The hardness of decoding random linear codes remains a strong assumption, but ensuring that code-based HE schemes are resilient against quantum attacks requires further investigation. Standardizing security reductions and proofs for homomorphic operations is essential for integrating code-based HE into post-quantum cryptographic standards \cite{AguilarMelchor2018}.

\item Integration with Secure Computation and Applications \\
HE is often used in conjunction with other cryptographic protocols, such as secure multi-party computation (MPC) and zero-knowledge proofs (ZKP). Research into how code-based HE can efficiently integrate with these techniques is still in its early stages. Potential applications include privacy-preserving blockchain technologies, federated learning, and secure genomic data processing \cite{Persichetti2020}.
\end{enumerate}

\bibliography{MAIN}
\bibliographystyle{plain}
\end{document}